\documentclass[11pt,a4paper]{article}
\usepackage{jheppub}
\usepackage{gensymb}
\usepackage{ae} 
\usepackage{aecompl}
\usepackage{bm} 
\usepackage{slashed}
\usepackage[usenames,dvipsnames]{color}

\usepackage{amsmath}
\usepackage{amssymb}
\usepackage{epsfig}

\usepackage[utf8]{inputenc}

\usepackage{fontenc}
\usepackage{graphicx}
\usepackage{float}
\usepackage[lofdepth,lotdepth,caption=false]{subfig}
\usepackage{color}
\usepackage{booktabs}
\usepackage{tabularx}
\usepackage{multirow}
\usepackage{longtable}
\usepackage{dcolumn}
\usepackage{rotating}%

\newcommand{\bi}{\begin{itemize}}
\newcommand{\ei}{\end{itemize}}

\def\beq{\begin{equation}}
\def\eeq{\end{equation}}

\newcommand{\bea}{\begin{eqnarray}}
\newcommand{\eea}{\end{eqnarray}}

\newcommand{\pme}{P({\nu_{\mu} \rightarrow \nu_{e}})}

\newcommand{\pmebar}{P_{\bar{\nu}_{\mu} \rightarrow \bar{\nu}_{e}}}
\newcommand{\nue}{\nu_{\mu} \rightarrow \nu_{e}}

\newcommand{\numu}{{\nu_{\mu}} \rightarrow \nu_{\mu}}

\newcommand{\ta}{\theta_{12}}
\newcommand{\tb}{\theta_{13}}
\newcommand{\tc}{\theta_{23}}
\newcommand{\dcp}{\delta_{13}}
\newcommand{\ldm}{\Delta m_{31}^2}
\newcommand{\sdm}{\Delta m_{21}^2}
\newcommand{\chisq}{\Delta \chi^{2}}
\newcommand{\dpme}{\Delta P_{\mu e}}
\newcommand{\ie}{{\it i.e.}}
\newcommand{\eg}{{\it e.g.}}
\newcommand{\etc}{{\it etc.}}
\newcommand{\nova}{{NO$\nu$A}}
\newcommand{\aee}{a_{ee}}
\newcommand{\amm}{a_{\mu\mu}}
\newcommand{\att}{a_{\tau\tau}}
\newcommand{\aem}{a_{e\mu}}
\newcommand{\aet}{a_{e\tau}}
\newcommand{\amt}{a_{\mu\tau}}
\newcommand{\phem}{\varphi_{e\mu}}
\newcommand{\phet}{\varphi_{e\tau}}
\newcommand{\phmt}{\varphi_{\mu\tau}}

\preprint{CTPU-PTC-22-15}
\title{
Investigating Lorentz Invariance Violation with the long baseline experiment P2O
}

  \author[a,1]{Nishat Fiza}
 \author[b,c,2]{Nafis Rezwan Khan Chowdhury}    
 \author[d,3]{Mehedi Masud}
 
\affiliation[a]{Department of Physical Sciences, IISER Mohali, Knowledge City, SAS Nagar, Mohali - 140306, Punjab, India}  
\affiliation[b]{Institut de F\'{i}sica Corpuscular (CSIC-Universitat de Val\`{e}ncia), Parc Cientific de la UV\\
C/ Catedratico Jos\'e Beltr\'an, 2, E-46980 Paterna (Val\`{e}ncia), Spain}
\affiliation[c]{Department of Physics and Astronomy, University of Utah\\
115 S. 1400 E., Salt Lake City, Utah 84112}
\affiliation[d]{Center for Theoretical Physics of the Universe, Institute for Basic Science (IBS), Daejeon 34126, Korea}
 \emailAdd{ph15039@iisermohali.ac.in} 
 \emailAdd{nafis.chowdhury@ific.uv.es}
\emailAdd{masud@ibs.re.kr}        
\abstract{
One of the basic propositions of quantum field theory is Lorentz invariance. 
The spontaneous breaking of Lorentz symmetry at a high energy scale can be studied at low energy extensions like the Standard model in a model-independent way through  
 effective field theory (EFT). 
The present and future Long-baseline neutrino experiments can give a scope to observe such a  Planck-suppressed physics of Lorentz invariance violation (LIV). 
A proposed long baseline experiment, Protvino to ORCA (dubbed "P2O") with a baseline of 2595 km, is expected to provide good sensitivities to unresolved issues, especially neutrino mass ordering. 
P2O can offer good statistics even with a moderate beam power and runtime, owing to the very large ($\sim 6$ Mt) detector volume at KM3NeT/ ORCA. 
Here we discuss in detail, how the individual LIV parameters affect neutrino oscillations at  P2O and DUNE baselines at the level of probability and derive analytical expressions to understand interesting degeneracies and other features. 
We estimate $\Delta \chi^{2}$ sensitivities to the LIV parameters, analyzing their correlations among 
each other, and also with the standard oscillation parameters. 
We calculate these results for P2O alone and also carry out a combined analysis of P2O with  DUNE. 
We point out crucial features in the sensitivity contours and explain them qualitatively with the help of the relevant probability expressions derived here. 
Finally we estimate constraints on the individual LIV parameters at $95\%$ confidence level (C.L.) intervals stemming from the combined analysis of P2O and DUNE datasets, and highlight the improvement over the existing 
constraints. 
We also find out that the additional degeneracy induced by the LIV parameter $a_{ee}$ 
around $-22 \times 10^{-23}$ GeV is lifted by the combined analysis at $95\%$ C.L.
}
\begin{document}  
\maketitle
\tableofcontents
%======================================
\section{Introduction}\label{sec:intro}                                                  %section I
%======================================
The phenomenon of neutrino oscillation which was first experimentally established more than twenty years back from the observations of  atmospheric and solar neutrinos~\cite{Fukuda:1998mi, Ahmad:2002jz} is one of the most transparent currently available portals into the rich physics beyond the standard model (BSM) of particle physics.  
In standard scenario neutrino oscillation is governed by six parameters, namely the three mixing angles ($\ta, \tb, \tc$); one Dirac CP phase ($\dcp$), and two mass-squared differences ($\sdm$, $\ldm$). 
So far $\ta$, $\tb$, $\sdm$ and the magnitude $|\ldm|$ have been measured with good precision from various neutrino experiments. 
One of the principal focus of the neutrino oscillation community is now on the measurement and implications of the values of the remaining parameters: the leptonic (Dirac) CP phase $\dcp$, the sign of $\ldm$ (denoting the correct neutrino mass ordering) and the octant of the mixing angle $\tc$. 
A value of $\dcp$ not equal to zero or $\pi$ would indicate CP violation in the lepton sector. This, in turn,  
can potentially shed light on the another fundamental puzzle, namely the baryon asymmetry of the universe~\cite{Sakharov:1967dj}. 
Resolution of the correct mass ordering and octant can help narrow down the plausible set of models explaining 
neutrino mass generation.

Presently running long-baseline neutrino oscillation experiments 
Tokai to Kamioka (T2K)~\cite{Abe:2013hdq} and NuMI Off-axis 
$\nu_e$ Appearance (\nova)~\cite{Ayres:2004js} are already giving us glimpses 
to the resolutions of the issues mentioned above. 
T2K data~\cite{Abe:2019vii} has ruled out CP conservation ($\dcp \simeq 0, \pi$) at 
$95\%$ confidence limit (C.L.). 
Irrespective of the mass ordering, at $99.73\%$ C.L. ($3\sigma$) T2K excludes $42\%$ of the 
entire parameter space for $\dcp$ (mostly around $+ \pi/2$), restricting the allowed region to 
roughly $\dcp \in [-\pi, 0.04\pi] \cup [0.89\pi, \pi]$. 
%Latest T2K results~\cite{Abe:2019vii} hint towards a HO value 
%for $\sin^{2}\theta_{23}$ = $0.53^{+0.03}_{-0.04}$ for both NO
%and IO.  
\nova\ data~\cite{Acero:2019ksn}, on the other hand indicates a slight preference for 
$\theta_{23}$ lying in the higher octant (HO) at a C.L. of $1.6\sigma$.
It also excludes most of the choices near 
$\dcp = \pi/2$ at a C.L. $\geqslant 3\sigma$ for inverted mass ordering (IO).
These measurements are expected to become more accurate as more data pour in.
Though the global analyses of neutrino data~\cite{deSalas:2020pgw, globalfit, Capozzi:2017ipn, Esteban:2018azc} shows an indication towards NO with $\tc$ possibly lying in the higher octant, the CP phase 
still has a large uncertainty. 

In near future, various other next-generation neutrino experiments with more sophisticated detection technologies are expected to start taking data. 
These experiments include, among others, 
Deep Underground Neutrino 
Experiment (DUNE)~\cite{Acciarri:2015uup, Abi:2020evt}, 
Tokai to Hyper-Kamiokande (T2HK)~\cite{Abe:2015zbg}, 
Tokai to Hyper-Kamiokande with a second detector in 
Korea (T2HKK)~\cite{Abe:2016ero},  European 
Spallation Source $\nu$ Super Beam 
(ESS$\nu$SB)~\cite{Baussan:2013zcy}, Jiangmen Undergound Neutrino Observatory (JUNO)~\cite{JUNO:2015zny}, Protvino to ORCA (P2O)~\cite{Akindinov:2019flp}. 
These experiments are expected to reach upto an unprecedented ($\sim$ a few percent) level of precision in 
measuring the oscillation parameters and hence are also susceptible to the presence of various possible new 
physics. 

%In the present work we explore the new physics effects of CPT violation by probing Lorentz Invariance Violation (LIV). 
CPT symmetry, one of the most sacred foundations in local relativistic quantum field theory, is based on the assumptions of the hermiticity of the hamiltonian, Lorentz invariance and local commutativity. 
Since an interacting theory with CPT violation also breaks Lorentz invariance~\cite{Greenberg:2002uu}, one 
widely used strategy to probe CPT violation is to analyze the associated Lorentz invariance violation (LIV). 
Spontaneous breakdown of Lorentz invariance may occur in theories of quantum gravity (in string theory, for \eg) at Planck scale ($M_{P} \sim 10^{19}$ GeV), forcing a Lorentz tensor field to acquire a non-zero vacuum expectation value, thus selecting a preferred spacetime direction~\cite{Kostelecky:1988zi, Kostelecky:1989jp, Kostelecky:1991ak, Kostelecky:1994rn, Kostelecky:1995qk}. 
It has been shown in literature that the Standard Model (SM) of particle physics can
be extended to construct a low energy effective field theory (EFT), namely 
Standard Model Extension (SME)~\cite{Colladay:1996iz, Colladay:1998fq, Kostelecky:2003fs} that includes such Lorentz invariance violating effects, suppressed by $M_{P}$. 
Neutrino oscillation by virtue of its interferometric nature, can probe such LIV effects at SME, thereby 
offering us a probe to the Planck scale physics. 

A broad range of experimental parameters such as neutrino-beam flavor composition, length, direction, and energy, as well as detector techniques provide different and often complementary sensitivities to the many higher dimensional operators characterizing LIV at accessible range of energies. 
Indeed, constraints on LIV parameters of SME have been obtained analysing the data from several neutrino experiments, - LSND~\cite{LSND:2005oop}, MINOS~\cite{MINOS:2008fnv, MINOS:2010kat},  MiniBooNE~\cite{MiniBooNE:2011pix}, Double Chooz~\cite{DoubleChooz:2012eiq}, Super-Kamiokande~\cite{Super-Kamiokande:2014exs}, T2K~\cite{T2K:2017ega}, IceCube~\cite{IceCube:2017qyp}.  
Outside the experimental collaboration also, there exist studies to explore LIV and CPT-violation ,- 
for \eg, in long-baseline accelerator 
neutrinos~\cite{Dighe:2008bu, Barenboim:2009ts, Rebel:2013vc, deGouvea:2017yvn, Barenboim:2017ewj, Barenboim:2018ctx, Majhi:2019tfi, KumarAgarwalla:2019gdj, Rahaman:2021leu}, 
short-baseline reactor antineutrinos~\cite{Giunti:2010zs}, atmospheric neutrinos~\cite{Datta:2003dg, Chatterjee:2014oda, Koranga:2014dua, Sahoo:2021dit}, 
solar neutrinos~\cite{Diaz:2016fqd}, and high-energy astrophysical neutrinos~\cite{Hooper:2005jp, Tomar:2015fha, Liao:2017yuy}. 
Recently, the authors of \cite{Lin:2021cst} have explored higher dimensional LIV parameters in the context of muon g-2 measurements by analysing available oscillation data for \nova\ and T2K.
For a comprehensive list of constraints on all the LIV parameters collected together we refer the readers to reference \cite{Kostelecky:2008ts}. 

The proposed P2O experiment~\cite{zaborov_talk, KM3Net:2016zxf, Zaborov:2018whl, Akindinov:2019flp} will have a baseline extending approximately 2595 km from the Protvino accelerator complex to the 
ORCA/KM3NET detector at the Mediterranean, - both of which are already existing. 
P2O baseline is most sensitive to first $\nue$ oscillation maxima around 4-5 GeV. 
Neutrino interaction around this energy is dominated by Deep Inelastic Scattering which is relatively 
well described theoretically, compared to, for \eg, 2-2.5 GeV (for DUNE) where resonant interactions and 
nuclear effects can potentially impact the measurements more significantly~\cite{Coloma:2013rqa, Mosel:2013fxa, Alvarez-Ruso:2014bla, Benhar:2015wva, NuSTEC:2017hzk, Nagu:2019fvi}. 
Such a very long baseline and relatively higher energy of the oscillation maxima gives P2O an excellent  
 level of sensitivity, especially towards neutrino mass ordering. 
 As has been illustrated in reference \cite{Singha:2021jkn}, the P2O baseline is 
  favourable to determine mass hierarchy also due to the much less interference by 
 the hierarchy-CP phase degeneracy. 
The very large detector volume of 6 Mt at ORCA will allow to detect thousands of neutrino events per year 
even with a very large baseline and a moderate beam power, - subsequently offering 
sensitivities to neutrino mass ordering, CP violation and $\tc$-octant that are competitive with the current and upcoming long-baseline neutrino experiments\footnote{P2O in its nominal configuration with a 90 kW beam, can resolve mass 
ordering with $\gtrsim 6\sigma$ sensitivity in 5 years of running, and also has a projected sensitivity of more than $3\sigma$ to $\tc$-octant with 3 years of running. With a 450 kW beam, it can offer $2\sigma$ sensitivity to $\dcp$ after 3 years of operation.}~\cite{Choubey:2018rnl, Akindinov:2019flp}. 
Recently it has been proposed that it is also possible to reach unprecedented sensitivity to leptonic CP violation at P2O using 
tagged neutrino beams by utilizing the kinematics of neutrino production in accelerators and recent advances in silicon particle detector technology~\cite{Perrin-Terrin:2021jtl}.
In recent years, there has been some interests in estimating new physics capabilities of P2O. 
Reference \cite{Kaur:2021rau} discussed the sensitivity reach of P2O to Non-unitarity of the leptonic mixing matrix and 
also estimated how it will affect the standard physics searches. 
The authors of \cite{Singha:2021jkn} discussed about the possible optimization of P2O in order to explore 
non-standard neutrino interactions. 
In the present work, we analyze the capabilities of P2O to probe violations of Lorentz invariance and CPT symmetry to  
estimate the constraints that can be put on these new physics parameters. 

The present manuscript is organised as follows. 
In Sec,\ \ref{sec:LIV_theory} we briefly describe the formalism of LIV. 
In Sec.\ \ref{sec:prob_LIV} we discuss in detail the probability expressions in presence of 
LIV parameters and provide a thorough analysis of the changes induced by each LIV 
parameter by means of heatplots. 
Sec.\ \ref{sec:analysis} describes the simulation procedures followed in this work. 
Secs.\ \ref{sec:LIV_corr} and \ref{sec:LIV_std_deg} illustrate the $\chisq$ sensitivity results 
showing the correlations of LIV parameters among themselves and with the standard oscillation parameters $\dcp$ and $\tc$. 
Sec.\ \ref{sec:bound} shows our final results as the constraints on LIV parameters obtained from this work, followed by the summary and conclusion in Sec.\ \ref{sec:summary}.

%which can heavily mess up the interpretation of the measured data. 
%Indeed the indications by global analysis of actual neutrino data can give a completely different neutrino mass ordering~\cite{Kelly:2020fkv}, $\tc$-octant~\cite{}, as well as leptonic CP violation~\cite{}.

%======================================
\section{Theoretical background}     
\label{sec:LIV_theory}                                             %section II
%======================================
We follow the widely used formalism of introducing Planck-suppressed CPT/Lorentz invariance 
violating effect to write a Lagrangian for the Standard Model Extension (SME), as 
developed in \cite{Colladay:1996iz, Colladay:1998fq, Kostelecky:2000mm, Kostelecky:2003cr, Kostelecky:2003fs, Diaz:2009qk, Kostelecky:2011gq}. The Lagrangian relevant for neutrino propagation in SME is then given by,
\begin{equation}\label{eq:lag}
\mathcal{L} = \frac{1}{2}\bar{\Psi}({i\slashed{\partial}} 
- M + \hat{\mathcal{Q}})\Psi + h.c.,
\end{equation}
where $\Psi$ is the spinor containing the neutrino fields. 
The first two terms inside the parentheses are the usual kinetic and the mass terms in 
the SM Lagrangian while the LIV effect has been incorporated by the operator $\hat{\mathcal{Q}}$. 
The Lorentz invariance violating term, which is suppressed by Planck-mass scale $M_{P}$ can be written in terms of the basis of the usual gamma-matrix algebra. 
Considering only renormalizable and only the CPT-violating LIV terms, one can write 
the LIV Lagrangian from Eq.\ \ref{eq:lag} in terms of vector and pseudovectors~\cite{Colladay:1996iz},
\begin{align}\label{eq:lag_liv}
\mathcal{L}_{\text{LIV}} \supset -\frac{1}{2}\left[ a^{\mu}_{\alpha\beta}\bar{\psi}_{\alpha}\gamma_{\mu}\psi_{\beta} + b^{\mu}_{\alpha\beta}\bar{\psi}_{\alpha}\gamma_{5}\gamma_{\mu}\psi_{\beta} 
%- i  c_{\alpha\beta}^{\mu\nu}   \bar{\psi}_{\alpha}\gamma_{\mu}\partial_\nu\psi_{\beta}
%- i d_{\alpha\beta}^{\mu\nu}   \bar{\psi}_{\alpha}\gamma_5\gamma_{\mu}\partial_\nu\psi_{\beta}
\right],
\end{align}
where $a_{\mu}, b_{\mu}$ are constant hermitian matrices and are in general combinations of tensor expectations, mass parameter and coefficients 
arising from the decomposition of gamma matrices. 
We focus on the following CPT-violating LIV parameter that is relevant in the context of the 
propagation of left handed 
neutrinos,
%========
\begin{align}
(a_{L})^{\mu}_{\alpha\beta} = (a + b)^{\mu}_{\alpha\beta}.
\end{align}
%========
Since our focus is on the isotropic component of the LIV  
terms, we will make the Lorentz indices zero. 
To further simplify our notation we will henceforth denote
the parameter $(a_{L})^{0}_{\alpha\beta}$ as $a_{\alpha\beta}$\footnote{
The presence of LIV makes it necessary to report the LIV bounds in a specific frame to conveniently compare the results from various experiments. Following the widely used practice in literature, the LIV coefficients used in our analysis are defined in the Sun-centered celestial equatorial frame. The Z direction points north along the earth's rotational axis, X direction points towards the vernal equinox, while the Y direction completes the right-handed coordinate system~\cite{Kostelecky:2003cr}. Observations performed in any other inertial frame of reference can be related to that in this Sun-centered frame via Lorentz transformations. We refer the reader to reference \cite{Kostelecky:2002hh} for more details on LIV-related measurements performed in other frames of reference and how they can be related to the standard Sun-centered celestial frame of reference.}.

Using spinor redefinitions to get rid of the non-trivial time derivatives in the Lorentz invariance violating Lagrangian in Eq.\ \ref{eq:lag} and carrying out some lengthy algebra with the resulting modified Dirac equation one can derive the Lorentz invariance violating effective hamiltonian relevant for ultrarelativistic, left-handed neutrino propagation through matter~\cite{Diaz:2009qk, Kostelecky:2011gq, Diaz:2011ia}.
\begin{align}\label{eq:h_cpt}
&H \simeq 
\underbrace{
\frac{1}{2E}U
\left(\begin{array}{ccc} m_{1}^{2} & 0 & 0 \\ 0 & m_{2}^{2} & 0 \\ 0 & 0 & m_{3}^{2} \\\end{array}\right)U^{\dagger}
}_{H_{\text{vac}}} 
+
\underbrace{
 \sqrt{2}G_{F}N_{e}\left(\begin{array}{ccc} 1 && \\ &0& \\ &&0\\ \end{array}\right)
}_{H_{\text{mat}}}
+ \underbrace{
\left(
\begin{array}{ccc}
a_{ee} & a_{e\mu} & a_{e\tau} \\
a^*_{e\mu} & a_{\mu\mu} & a_{\mu\tau} \\
a^*_{e\tau} & a^*_{\mu\tau} & a_{\tau\tau}
\end{array}
\right)
}_{H_{\text{LIV}}} \nonumber \\
&= \frac{1}{2E}U
\left(\begin{array}{ccc} 0 & 0 & 0 \\ 0 & \sdm & 0 \\ 0 & 0 & \ldm \\\end{array}\right)U^{\dagger}
+ \sqrt{2}G_{F}N_{e}\left(\begin{array}{ccc} 1 && \\ &0& \\ &&0\\ \end{array}\right)
+ \left(
\begin{array}{ccc}
a_{ee}-a_{\tau\tau} & a_{e\mu} & a_{e\tau} \\
a^*_{e\mu} & a_{\mu\mu}-a_{\tau\tau} & a_{\mu\tau} \\
a^*_{e\tau} & a^*_{\mu\tau} & 0
\end{array}
\right).
\end{align}
The first term containing the usual leptonic mixing matrix $U$ and the neutrino mass eigenstates $m_{i} (i=1,2,3)$ is the standard vacuum hamiltonian. 
The second term, proportional to Fermi constant $G_{F}$ and electron density $N_{e}$ along the 
neutrino propagation, originates from standard charged-current coherent forward scattering of neutrinos with electrons in earth matter. 
The third term containing the LIV parameters $a_{\alpha\beta}$'s ($\alpha, \beta = e, \mu, \tau$) 
incorporates the effect of LIV (and also CPT-violation). 
The off-diagonal $a_{\alpha\beta}$'s ($\alpha \neq \beta$) are complex with a phase ($\varphi_{\alpha\beta}$) associated to them, while the diagonal parameters are real. 
As per convention, in the second line of Eq.\ \ref{eq:h_cpt}, a common term ($m_{1}^{2}$) has been subtracted from the diagonal elements in $H_{\text{vac}}$, and another 
common term $\att$ has been subtracted from the diagonal elements of $H_{\text{LIV}}$. 
Both of these subtractions have the effect of removing an overall phase factor, which 
will have no impact on the oscillation probabilities. 
This implies that neutrino oscillation can effectively probe only two of the three diagonal parameters in $H_{\text{LIV}}$. 
In our analysis, those two parameters are 
$\tilde{a}_{ee} = \aee-\att$ \& $\tilde{a}_{\mu \mu} = \amm-\att$, while the individual value of $\att$ 
cannot be probed by the oscillation experiment. 
Thus for simplicity we take $\att$ to be zero and thus $\tilde{a}_{ee} = \aee$, and $\tilde{a}_{\mu \mu} = \amm$. 
Note that, any one of the three diagonal LIV parameters can be chosen to be removed from the analysis in this way.

It is worthwhile to mention here that the physics of neutral current (NC) Nonstandard interaction (NSI) (usually denoteds by $\varepsilon_{\alpha\beta}$) that arises from neutrino mass models and 
introduces couplings between the neutrinos and the first generation fermions $e, u, d$, has an 
apparent similarity with the form of LIV hamiltonian, - thereby suggesting a mathematical mapping: $\varepsilon_{\alpha\beta} \leftrightarrow  a_{\alpha\beta}/\sqrt{2}G_{F}N_{e}$.  
But there is a crucial difference between these two different kinds of physics scenario as discussed in detail in reference \cite{Diaz:2015dxa}. 
NC NSI is proportional to the density along the neutrino trajectory and is thus very tiny for 
short-baseline neutrino experiments. LIV, on the other hand, is an intrinsic effect that is present even in the vacuum.

%==============================================
 \section{Impact of LIV parameters on probability}                         %Section IV
 \label{sec:prob_LIV}
 %================================================
 %--------------------------------------------
 \begin{figure}[h]
 \centering
 \includegraphics[scale=0.55]{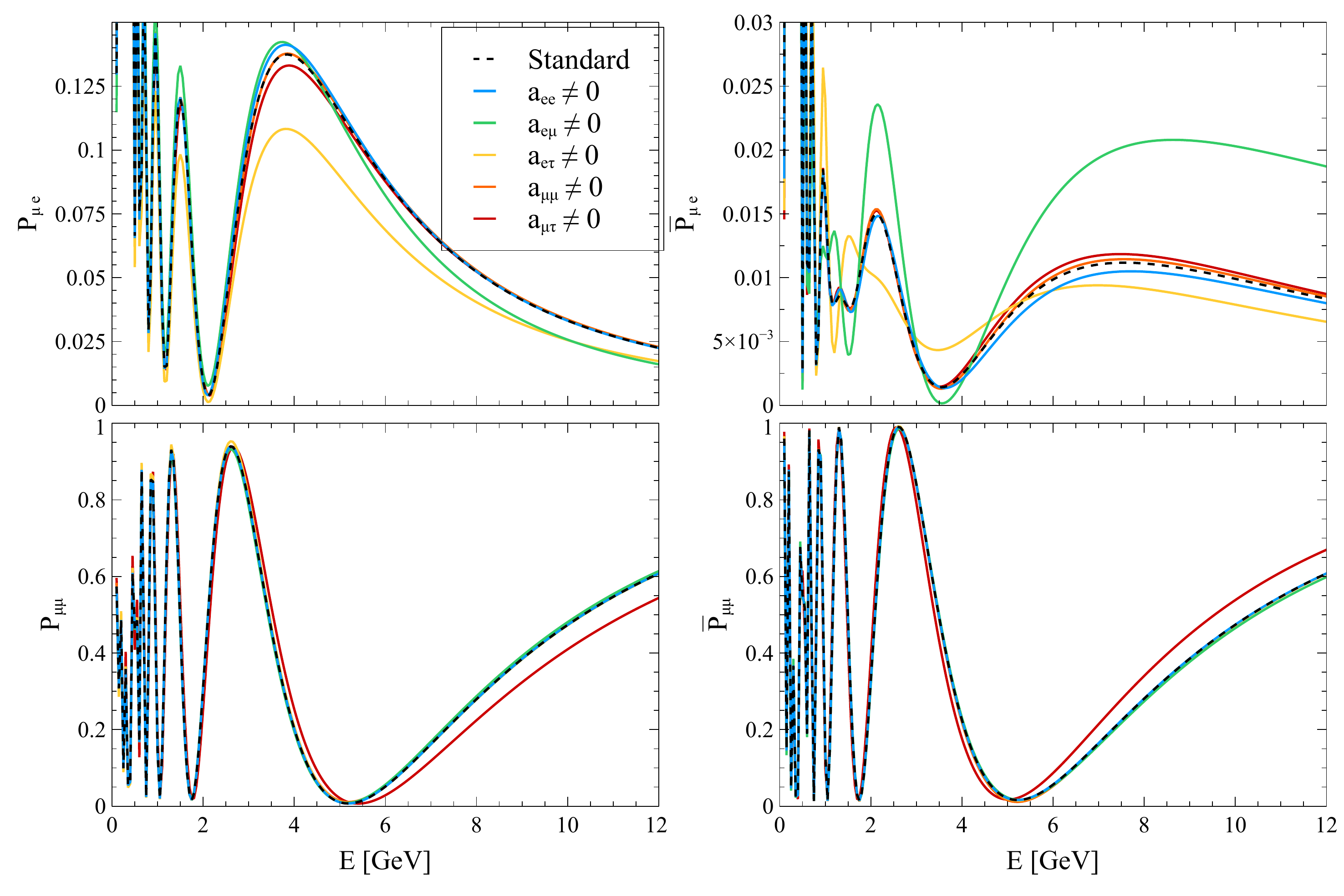}
 \caption{\footnotesize{The impact of individual LIV parameters on oscillation probability at the P2O baseline of 2595 km. The top (bottom) row shows the probability for appearance (disappearance) channel, while the left (right) column indicates the neutrino (anti-neutrino) mode.
 The black dotted curve shows the probability for the standard (no LIV) case, while the solid coloured curves 
 are for individual LIV parameters present.
 The non-zero values of the individual LIV parameters $a_{\alpha\beta}$ was taken as $5 \times 10^{-23}$ GeV, while the CP phase associated with the off-diagonal LIV parameters was taken as zero.}}
  \label{fig:prob}
 \end{figure}
%-----------------------------------------------
In this work we focus on $a_{\alpha\beta}$'s and we will now describe 
how they affect the oscillation probability expressions in various channels.
Since the main contribution comes from the $\nue$ oscillation channel, we 
discuss about how $\pme$ is affected by LIV. The most important LIV parameters 
impacting this channel are $\aem$ and $\aet$, and also to a lesser extent $\aee$. 
Following the similar approach as in reference \cite{Kikuchi:2008vq, Agarwalla:2016fkh,Masud:2018pig, KumarAgarwalla:2019gdj}, 
we can approximately write the $\nu_{\mu} \to \nu_{e}$ oscillation probability as the sum of the following three terms, 
\begin{align}\label{eq:pme_liv}
P_{\mu e} (\text{SI+LIV}) \simeq P_{\mu e}(\text{SI}) + P_{\mu e}(\aem) + P_{\mu e}(\aet), 
\end{align} 
where the first term on the right hand side is the probability term corresponding to standard interaction (SI) with earth matter, while the other
two terms come due to the presence of LIV parameters $\aem$ and $\aet$. 
The three terms on the right hand side can be shown to have the following forms.
%%%%%%%%%%%%%%%%%%%%
\begin{align}\label{eq:p_si}
P_{\mu e}(\text{SI}) \simeq X + Y\cos(\dcp + \Delta),
\end{align}
%%%%%%%%%%%%%%%%%%%%
%%%%%%%%%%%%%%%%%%%%%
\begin{align}
\label{eq:p_aem} 
&P_{\mu e}(a_{e\mu})  \nonumber \\
&\simeq \frac{8 |a_{e\mu}| E \Delta s_{13} \sin2\theta_{23} c_{23}\sin\Delta}{\ldm} 
\bigg[
-\sin\Delta \sin (\dcp+\phem) + \bigg(\frac{s^{2}_{23}}{c^{2}_{23}}\frac{\sin\Delta}{\Delta} +\cos\Delta\bigg)
\cos(\dcp + \phem) \bigg], \\
\label{eq:p_aet} 
&P_{\mu e}(a_{e\tau}) \nonumber \\
&\simeq \frac{8 |a_{e\tau}| E \Delta s_{13} \sin2\theta_{23} s_{23}\sin\Delta}{\ldm} 
\bigg[
\sin\Delta \sin (\dcp+\phet) + \bigg(\frac{\sin\Delta}{\Delta} - \cos\Delta\bigg) 
\cos(\dcp + \phet) \bigg]. 
\end{align}
%%%%%%%%%%%%%%%%%%%%%%
The different familiar terms appearing in Eqs.\ \ref{eq:p_si}, \ref{eq:p_aem}, \ref{eq:p_aet} 
are given in the following.
\begin{align}\label{eq:si_coeff}
&X = 4s_{13}^{2}c_{13}^{2} s_{23}^{2} \frac{\sin^{2}\big[(1-\hat{A})\Delta \big]}{(1-\hat{A})^{2}}; \qquad
Y = 8\alpha s_{12} c_{12} s_{23} c_{23} s_{13}c_{13} \frac{\sin \hat{A}\Delta}{\hat{A}}  \frac{\sin \big[(1-\hat{A})\Delta\big]}{1-\hat{A}}, \nonumber \\
&\hat{A} = \frac{2\sqrt{2}G_{F}N_{e}E}{\ldm}; \qquad  \Delta = \frac{\ldm L}{4E}; \qquad s_{ij} = \sin\theta_{ij}; \qquad c_{ij} = \cos\theta_{ij}; \qquad \alpha = \frac{\sdm}{\ldm}.
\end{align}
%In writing the expression for $P_{\mu e}(\text{SI})$ 
%in Eq.~\ref{eq:p_si}, we neglect the {\it{solar}} term 
%$\alpha^{2} \sin^{2} 2\theta_{12} c_{23}^{2} \frac{\sin^{2}\hat{A}\Delta}{\hat{A}^{2}}$.
%This is due to the fact that by considering the values 
%of the oscillation parameters as $\theta_{12} = 34.5^{\circ}, 
%\theta_{13} = 8.45^{\circ}, \theta_{23} = 47.7^{\circ}, 
%\sdm = 7.5 \times 10^{-5} \text{ eV}^{2}, 
%\ldm = 2.5 \times 10^{-3} \text{ eV}^{2}$ 
%(which are in agreement with~\cite{deSalas:2017kay,globalfit,Capozzi:2018ubv,Esteban:2018azc}), 
%we find that the {\it{solar}} term, being proportional to $\alpha^{2}$, 
%is roughly suppressed by 3 to 4 orders of magnitude 
%as compared to the other two terms as shown 
%in Eq.~\ref{eq:p_si}.
In presence of $\aee$, the replacement $\hat{A} \to \hat{A}[1+\aee/\sqrt{2}G_{F}N_{e}] \simeq \hat{A}+\aee/(2E/\ldm)$ has to be made.
In order to understand the impact of the LIV parameters, we first have a look at the oscillation probability at the P2O baseline of 2595 km. 
This was estimated numerically using the widely used 
General Long Baseline Experiment Simulator (GLoBES)~\cite{Huber:2004ka, Huber:2007ji} 
and the associated package \textit{snu.c}~\cite{Kopp:2006wp,Kopp:2007ne} with necessary modifications. 
We consider Normal mass ordering (NO) and take the following best fit values~\cite{deSalas:2020pgw} of the oscillation parameters: 
$\ta = 34.3^{\circ}, \tb = 8.58^{\circ}, \tc = 48.8^{\circ}, \dcp = -0.68\pi, \sdm = 7.5\times 10^{-5} \text{ eV}^{2}, 
\ldm = 2.5 \times 10^{-3} \text{ eV}^{2}$. 
We take one LIV parameter $a_{\alpha \beta}$ non-zero (fixed at the same numerical value of $5 \times 10^{-23}$ GeV, and the associated CP phase $\varphi_{\alpha\beta} = 0$) at a time to assess the role of individual LIV parameters in the probability level, and show the results in Fig.\ \ref{fig:prob}.

As expected, the appearance channel is most affected by the LIV parameters $\aem$ and $\aet$. 
Compared to the standard case (black dashed curve), $\aem$ increases the magnitude of $\pme$ while the presence of $\aet$ shows a depletion around the oscillation maxima of $4-5$ GeV. 
This is due to the fact that both the $\sin\dcp$ and $\cos\dcp$ terms within the square 
brackets of Eq.\ \ref{eq:p_aet} have the same sign (negative, thus decreasing $P_{\mu e}$), while there is 
a relative sign between two such terms in Eq.\ \ref{eq:p_aem}, - thus leading to a smaller 
enhancement of $P_{\mu e}$. 
The effects of $\aem$ and $\aet$ become qualitatively opposite for the $\pmebar$ channel. 
We also observe that $\aee$ increases or decreases the probabilities only mildly. 
The disappearance channel, on the other hand is impacted by only the parameters $\amm$ and $\amt$, - the changes induced by them being in the opposite direction for $\nu$ and $\bar{\nu}$-modes.

%%%%%%%%%%%%%%%%%%%%% 
  \begin{figure}[b]
 \centering
 \includegraphics[scale=0.5]{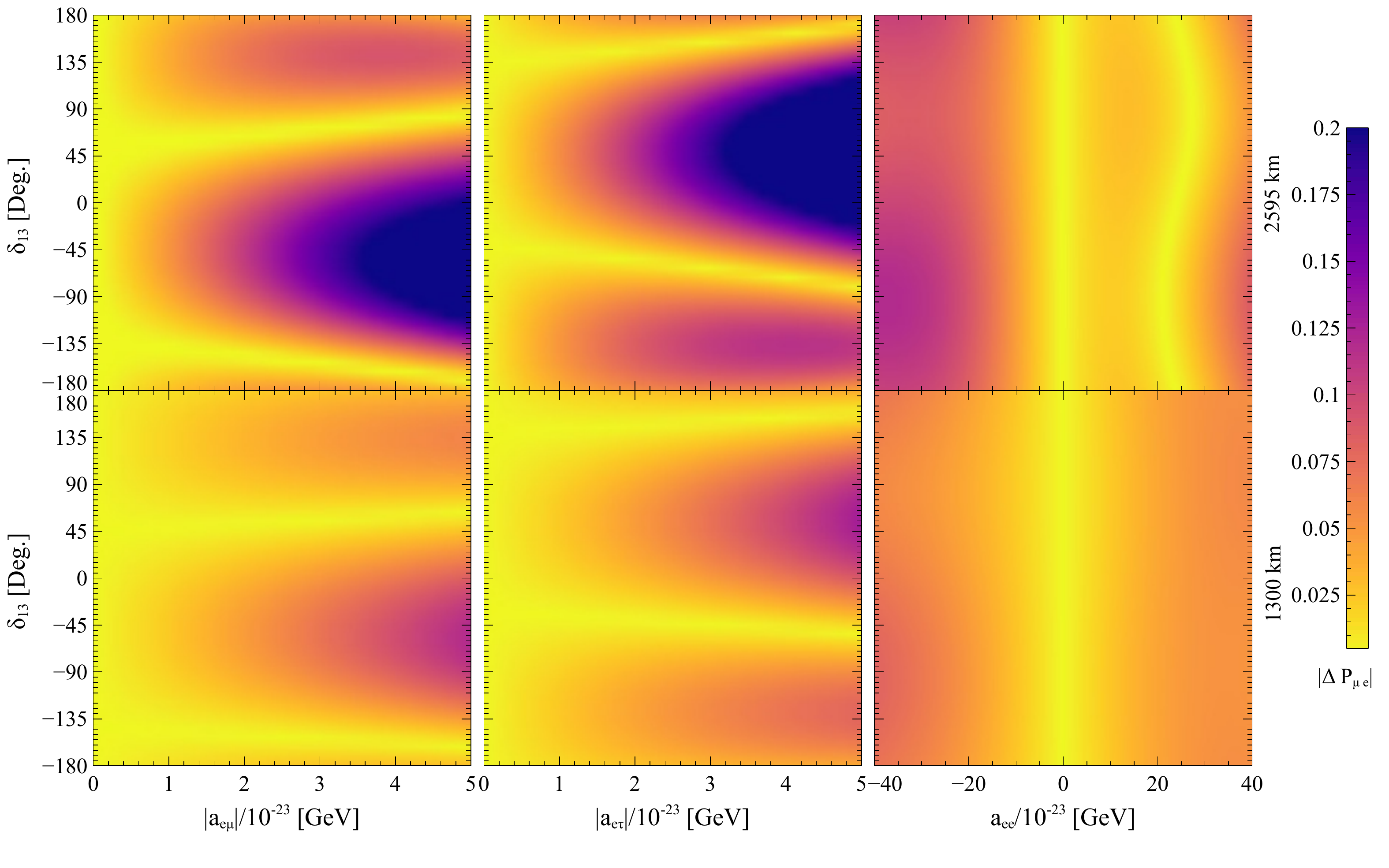}
 \caption{\footnotesize{We plot the heatplot for $|\dpme| = |P_{\mu e} (\text{SI+LIV}) - P_{\mu e} (\text{SI})|$ for the P2O baseline of 2595 km (top row) and for DUNE baseline of 1300 km 
 (bottom row). The plots were done for the respective values of energies roughly corresponding to 
 first oscillation maximum: 2.5 GeV for DUNE and 5 GeV for P2O. 
 The heatplot is shown as a function of the three LIV parameters ($|\aem|, |\aet|, \aee$) 
 along the horizontal axes and the standard (Dirac) CP phase $\dcp$.}}
  \label{fig:heatplot_dcp}
 \end{figure}
 %%%%%%%%%%%%%%%%%%%%%%%
The sensitivity to the LIV parameters depends on the change in 
 probability due to the presence of LIV: 
 \begin{equation}\label{eq:dpme}
 \Delta P_{\alpha\beta} = P_{\alpha\beta} (\text{SI+LIV}) - P_{\alpha\beta} (\text{SI}), 
 \qquad ( \alpha,\beta = e, \mu, \tau).
 \end{equation}
 In order to have an approximate idea about the physics behind the sensitivity estimates, 
 we focus on the dominant channel, \ie, the $\nue$ channel and the most relevant LIV parameters $\aem, \aet, \aee$. 
 
In Fig.\ \ref{fig:heatplot_dcp} we show by means of a heatplot, how the absolute difference $|\dpme|$ evolves with variation in the LIV parameters and the variation in the standard CP phase $\dcp$, for a fixed 
baseline and energy. 
In top (bottom) row, we consider the baseline 2595 (1300) km and approximate first oscillation maximum energy 5 (2.5) GeV for the P2O (DUNE) experiment. 
The light yellow end of the colour spectrum corresponds to lower $|\dpme|$ (\ie, more degeneracy between SI and LIV), 
while the darker shades indicate a higher impact of the corresponding LIV parameter, 
resulting in a higher value of $|\dpme|$. 

In all the heatplots we see that there is little to no change in the probability for very small values 
of the LIV parameter, which is consistent with our expectation. 
In presence of $\aem$ ($\aet$), we note the presence of a set of two degenerate (yellow) {\textit{branches}} appearing at two different values of $\dcp$. 
Interestingly these degeneracies remain present irrespective of the values of $|\aem|$ or 
$|\aet|$ and the degenerate regions are almost parallel to the LIV parameter axis. 
These features are more prominent for the P2O baseline than the DUNE baseline. 
On the other hand, in presence of $\aee$, P2O baseline shows an additional degeneracy 
approximately around $\aee \simeq 22 \times 10^{-23}$ GeV, but curiously this is absent for 
DUNE. 
 
For an analytical understanding of the various features, we use Eqs.\ \ref{eq:pme_liv},  \ref{eq:p_si}, 
\ref{eq:p_aem}, \ref{eq:p_aet} 
to express $|\dpme|$ in presence of $\aem$ or $\aet$ as the following. 
\begin{align}
\label{eq:dpme_aem}
\dpme(|\aem|) &\simeq
8 |\aem| \frac{\pi}{2} E  s_{13} \sin 2\tc c_{23}
\bigg[ -\sin\dcp + \frac{2}{\pi} \frac{s^{2}_{23}}{c^{2}_{23}} \cos\dcp \bigg], \\
\label{eq:dpme_aet}
\dpme(|\aet|) &\simeq
8 |\aet| \frac{\pi}{2} E  s_{13} \sin 2\tc s_{23}
\bigg[ \sin\dcp + \frac{2}{\pi} \cos\dcp \bigg].
\end{align}
Note that, since all the heatplots are generated corresponding to the first oscillation maximum, 
we put $\Delta = \ldm. L/E \simeq \pi/2$ in deriving Eqs.\ \ref{eq:dpme_aem} and \ref{eq:dpme_aet}. 
$\dpme(|\aem|)$ (or $\dpme(|\aet|)$) is directly proportional to $|\aem|$ (or 
$|\aet|$) respectively, - which clearly shows that for very small values of the LIV parameters, we get 
a degeneracy. 
For the pair of yellow degenerate {\textit{branches}} in the first two columns of Fig.\ \ref{fig:heatplot_dcp}, the quantities inside the square brackets of Eqs.\ \ref{eq:dpme_aem} and \ref{eq:dpme_aet} need to vanish, and the solutions are independent of the value of the LIV parameter and the baseline. 
For the case of $|\aem|$, this condition becomes,
%%%%%%%%%%%%%%%%%%%%
\begin{equation}\label{eq:zero_dpme_aem}
\sin\dcp = \frac{2}{\pi} \frac{s^{2}_{23}}{c^{2}_{23}} \cos\dcp.
\end{equation}
%%%%%%%%%%%%%%%%%%%%
Putting $\tc = 48.8^{\circ}$, the solutions are $\dcp \simeq 39^{\circ}, -141^{\circ}$.
It is clear from the $(s_{23}/c_{23})^{2}$ factor that for $\tc$ lying in the higher octant, 
first solution for $\dcp$ will move (mildly) closer to $\pi/4$, making the second solution move 
towards $-3\pi/4$. 
For the case of $|\aet|$, using Eq.\ \ref{eq:dpme_aet} the degenerate condition 
translates to, 
%%%%%%%%%%%%%%
 \begin{equation}\label{eq:zero_dpme_aet}
\sin\dcp = -\frac{2}{\pi} \cos\dcp, 
\end{equation} 
%%%%%%%%%%%%%%
the solutions of which are given by roughly $\dcp \simeq -33^{\circ}, 147^{\circ}$. 
We note that the solutions for $\dcp$ for Eqs.\ \ref{eq:zero_dpme_aem} and \ref{eq:zero_dpme_aet} for the locations of degeneracies approximately differ by a sign 
(as long as $\tc$ does not lie too far from the maximal value of $\pi/4$), or equivalently they differ by a $\pm \pi/2$ phase-shift. 
These locations of degeneracies and the shift of the solutions for $|\aem|$ and $|\aet|$ are consistent with Fig.\ \ref{fig:heatplot_dcp}.   
The slight {\textit{slanting}} nature of the degenerate branches with increase in $|\aem|$ 
originates due to subdominant higher order terms, which we have not considered in our 
simplified analysis. 
In Fig.\ \ref{fig:heatplot_dcp}, we note that deviation from the standard case 
happens more quickly when the CP phase $\dcp \in [-\pi/2, 0]$ (for $|\aem|$) and 
$\dcp \in [0, \pi/2]$ (for $|\aet|$), - manifested by the presence of darker patches around $|\aem|$  or $|\aet|$ $\gtrsim 2 \times 10^{-23}$ GeV. 
These two separate quadrants for $\dcp$ originate due to the presence of the relative sign between the $\sin\dcp$ and $\cos\dcp$ terms inside 
the square brackets in Eqs.\ \ref{eq:dpme_aem} and \ref{eq:dpme_aet}. 
The proportionality of Eqs.\ \ref{eq:dpme_aem} and \ref{eq:dpme_aet} with energy suggests 
that the features are quantitatively more prominent for P2O than DUNE, since the peak 
energy corresponding to the former is twice the latter (5 GeV, as compared to 2.5 GeV for DUNE).  
To understand the features induced by the presence of $\aee$, we deduce the corresponding 
probability difference as follows (using Eq.\ \ref{eq:p_si} and replacing $\hat{A} \to \hat{A}[1+\aee/\sqrt{2}G_{F}N_{e}]$ to account for $\aee$).
 %%%%%%%%%%%%%%%%%%%
 \begin{equation}
\label{eq:dpme_aee}
\dpme(\aee) \simeq
4s_{13}^{2}c_{13}^{2} s_{23}^{2}\Bigg\{
\frac{\sin^{2}\big[1-\hat{A}(1+\aee/\sqrt{2}G_{F}N_{e})\big]\Delta}
{\big[1-\hat{A}(1+\aee/\sqrt{2}G_{F}N_{e})\big]^{2}}
- \frac{\sin^{2}\big[1-\hat{A}\big]\Delta}
{\big[1-\hat{A}\big]^{2}}
\Bigg\}
+ \cos\dcp\text{-term}.
\end{equation}
 %%%%%%%%%%%%%%%%%%
 The $\cos\dcp$-term containing Y from Eq.\ \ref{eq:p_si} is suppressed by a factor 
 $\alpha$ ( = $\sdm/\ldm \sim 10^{-2}$) compared to the first term in Eq.\  \ref{eq:dpme_aee}. 
We neglect this term for simplicity. 
Thus the degeneracy condition ($\dpme(\aee) \simeq 0$) in presence of $\aee$ can be simplified to the following 
equation. 
 %%%%%%%%%%%%%%
 \begin{align}\label{eq:zero_dpme_aee}
 \Bigg[
 \underbrace{
\frac{\sin\big[1-\hat{A}(1+\hat{a}_{ee})\big]\Delta}
{1-\hat{A}(1+\hat{a}_{ee})}
- \frac{\sin\big[1-\hat{A}\big]\Delta}
{1-\hat{A}} 
}_{\text{I}_{-}}
\Bigg]
\times \Bigg[
\underbrace{
\frac{\sin\big[1-\hat{A}(1+\hat{a}_{ee})\big]\Delta}
{1-\hat{A}(1+\hat{a}_{ee})}
+\frac{\sin\big[1-\hat{A}\big]\Delta}
{1-\hat{A}} 
}_{\text{I}_{+}}
\Bigg]
= 0,
\end{align} 
%%%%%%%%%%%%%%%%%%%%%%
where $\hat{a}_{ee} = \aee/\sqrt{2}G_{F}N_{e}$. 
%%%%%%%%%%%%%%%%%%%%% 
  \begin{figure}[htb]
 \centering
 \includegraphics[scale=0.6]{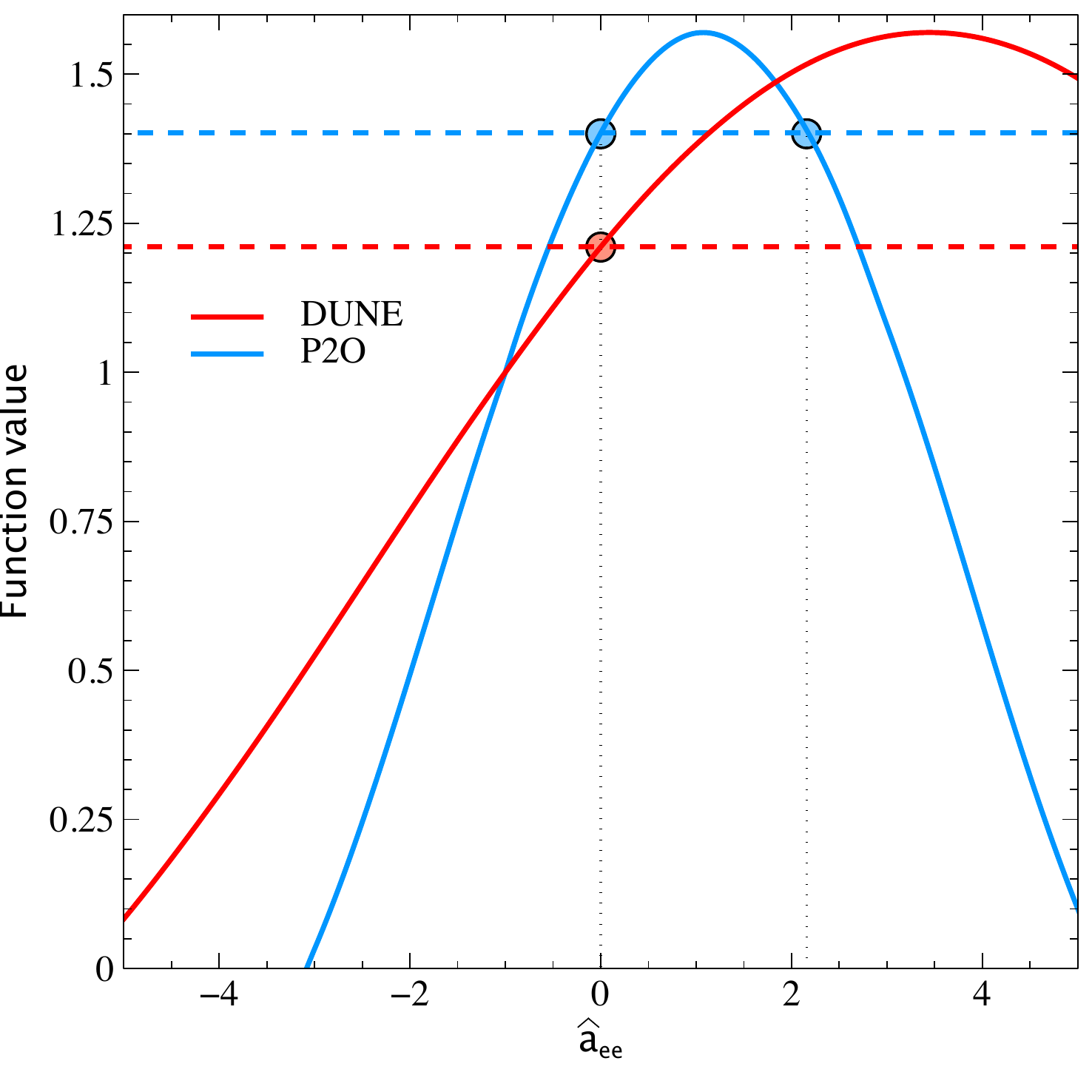}
 \caption{\footnotesize{The two terms in $\text{I}_{-}$ from Eq.\ \ref{eq:zero_dpme_aee} 
 are plotted for both DUNE (red) and P2O (blue) as functions of the parameter 
 $\hat{a}_{ee} = \aee/\sqrt{2}G_{F}N_{e}$. 
 The solid curve is the first term $\big(\frac{\sin[1-\hat{A}(1+\hat{a}_{ee})]\Delta}
{1-\hat{A}(1+\hat{a}_{ee})}\big)$, while 
 the dashed curve is the second term $\big(\frac{\sin[1-\hat{A}]\Delta}
{1-\hat{A}}\big)$. 
 The small coloured circles show the locations of solutions where the two terms intersect.}}
  \label{fig:analyze_aee}
 \end{figure}
 %%%%%%%%%%%%%%%%%%%%%%%
It is easy to see that $\text{I}_{+}$ cannot be zero, and for $\text{I}_{-}$ to vanish we can immediately identify $\aee = 0$ as the trivial 
solution. To examine the possibility of further degeneracies, we note the following.
%%%%%%%%%%%%%%%%%%%%
\begin{align}\label{eq:zero_dpme_aee_1} 
&\Delta \simeq \pi/2; \qquad (\text{for both P2O and DUNE}) \nonumber \\
&\hat{A} \simeq \frac{2\sqrt{2}G_{F}N_{e}E}{\ldm} 
\simeq 0.03 \times \rho[\text{g.cm}^{-3}] \times \text{E[GeV]} \simeq
\begin{cases}
0.225, \quad (\text{for DUNE}, \rho \simeq 3, E \simeq 2.5) \\
0.448. \quad (\text{for P2O}, \rho \simeq 3.2, E \simeq 5).
\end{cases}
\end{align} 
%%%%%%%%%%%%%%% %%%%%
To find other solutions when ${\text{I}_{-}} = 0$, we plot the two terms in ${\text{I}_{-}}$ 
for both DUNE and P2O as a function of the parameter $\hat{a}_{ee}$ in Fig.\ \ref{fig:analyze_aee}. 
 %%%%%%%%%%%%%%%%%%%%% 
  \begin{figure}[h]
 \centering
 \includegraphics[scale=0.5]{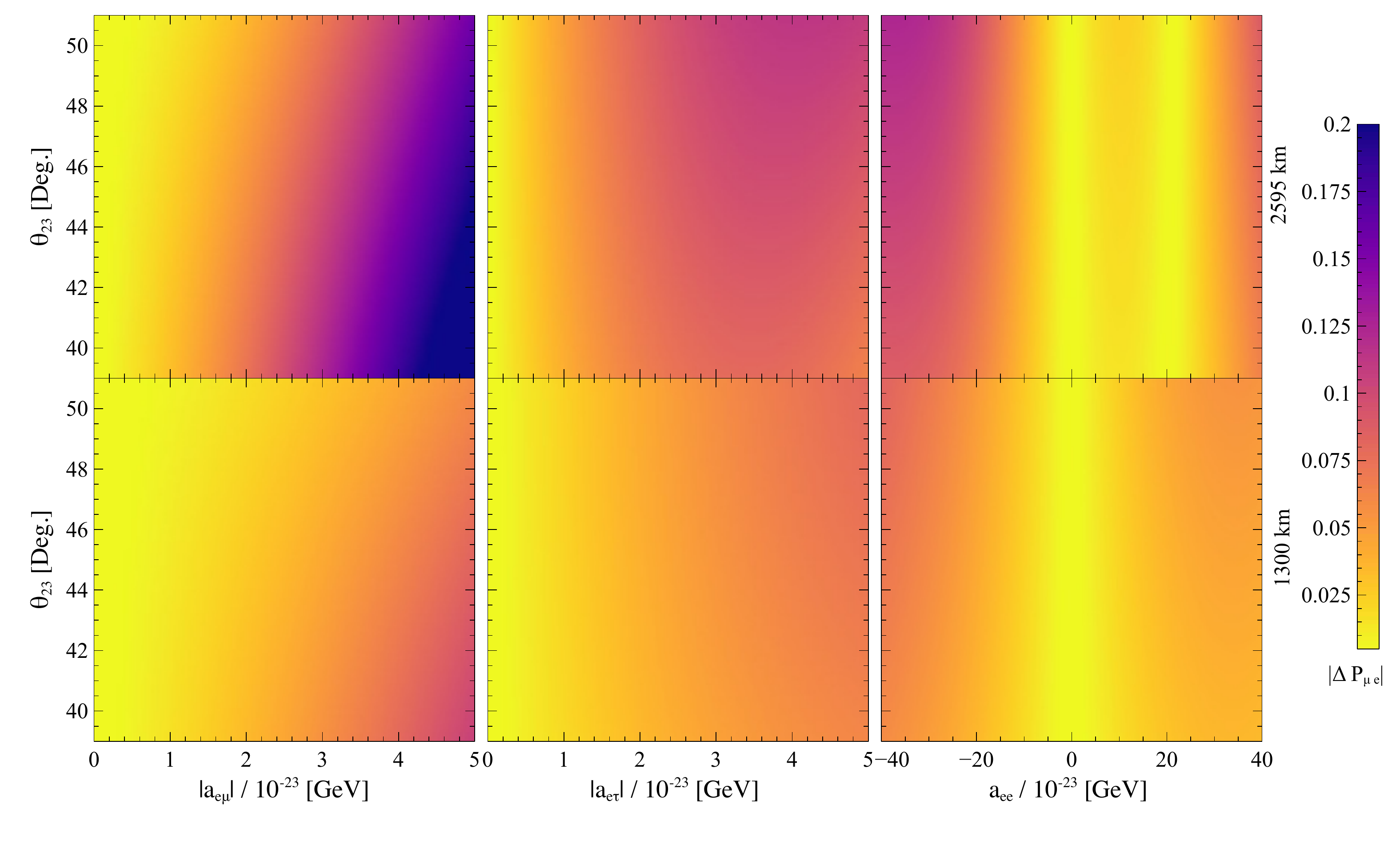}
 \caption{\footnotesize{Similar to Fig.\ \ref{fig:heatplot_dcp} but shown as a function of $\tc$ 
 with fixed $\dcp$.}}
  \label{fig:heatplot_th23}
 \end{figure}
 %%%%%%%%%%%%%%%%%%%%%%%
The first term is an oscillating function of $\hat{a}_{ee}$, while the second term is a constant. 
For DUNE, having a lower baseline and energy, the {\textit{sine}} function (red solid) oscillates slowly 
and has only the trivial solution in the range shown. 
Corresponding {\textit{sine}} function for P2O (blue solid) oscillates faster, given the larger 
baseline and energy, and thus can have a second (non-trivial) solution at a reasonably smaller positive value of $\hat{a}_{ee} \simeq 2.2$, which translates to ${a}_{ee} = 2.2 \sqrt{2}G_{F}N_{e} \simeq 24.8 \times 10^{-23}$ GeV. 
This is almost exactly the location of the second degeneracy in the top right panel of Fig.\ \ref{fig:heatplot_dcp}. 
The mild dependence of this degeneracy brach on the CP phase $\dcp$ arises from the 
$\cos\dcp$-term in Eq.\ \ref{eq:dpme_aee} which we have neglected for simplicity.

In Fig.\ \ref{fig:heatplot_th23}, we show the heatplot for $|\dpme|$ in the parameter space of 
$\tc$ and one LIV parameter ($|\aem|, |\aet|, \aee$), for a fixed CP phase $\dcp = -0.68\pi$.
Comparing the first and the middle columns, we see that $\aem$ has a slightly bigger impact 
than $\aet$.  
Moreover, presence of $\aem$ induces more deviation at lower octant (LO), while that of 
$\aet$ is apparent at higher octant (HO). 
If we look at the analytical expressions for $\dpme$ in Eqs.\ \ref{eq:dpme_aem} and \ref{eq:dpme_aet}, this octant dependendence originates due to the overall factor $c_{23}$ in 
presence of $\aem$ and $s_{23}$ in 
presence of $\aet$ (note that the factor $\sin2\tc$ in those equations are octant-independent). 
In the third column of Fig.\ \ref{fig:heatplot_th23}, $\aee$ again gives rise to additional degeneracy for P2O, which has already been explained above with regard to Fig.\  \ref{fig:heatplot_dcp}.
%==============================
\section{Simulation details}                                                             %section III
\label{sec:analysis}
%==============================
We simulate the long baseline neutrino experiments DUNE and P2O using GLoBES~\cite{Huber:2004ka, Huber:2007ji} and use the add-on {\it{snu.c}}~\cite{Kopp:2006wp,Kopp:2007ne} to implement the physics of LIV.
DUNE is a 1300 km long baseline experiment from the accelerator at the site of FermiLab 
to the site employing a liquid argon far detector (FD) of 40 kt fiducial mass at South Dakota. 
 The experiment is capable of using a proton beam of power 1.07 MW and of running 3.5 years each on $\nu$ and 
$\bar{\nu}$ mode (resulting in a total exposure of roughly 300 kt.MW.yr corresponding to total 
$1.47 \times 10^{21}$ protons on target or POT). 
The flux, cross-sections, migration matrices for energy reconstruction, efficiencies \etc\ 
were implemented according to the official configuration files~\cite{Alion:2016uaj} provided by the DUNE collaboration for its simulation.

P2O (Protvino to ORCA) is a proposed long baseline neutrino experiment with a baseline 
of nearly 2595 km from the Protvino accelerator complex, situated at $100$ km south of Moscow to the site of ORCA (Oscillation Research with Cosmics in the Abyss), hosting 6 MT 
Cerenkov detector located 40 km 
off the coast in South France, at a mooring depth of 2450 m in the Mediterranean sea. 
ORCA is the low energy component of the KM3NeT Consortium \cite{Adrian-Martinez:2016fdl},   
with a primary goal of studying atmospheric neutrino oscillations in the energy range of 3 to 100 GeV in order to determine the neutrino mass ordering. 
Currently, 10 lines (\ie, detection units) of the ORCA detector are live and taking data.  
A full ORCA detector is expected to have 115 lines and foresees completion in subsequent phases around 2025~\cite{Margiotta:2022kid}. 
Construction of the neutrino beamline and relevant upgradation of the accelerator for the P2O experiment is expected to be completed in a few years. 
Assuming a favorable geopolitical situation and available funding, the P2O project in its nominal configuration might be realised during the next decade~\cite{brunner_eppsu}.
We simulate the nominal configuration\footnote{There are proposals for using an upgraded proton beam with 450 kW power and also to use the Super-ORCA detector with denser geometry, 
lower energy thresholds and better flavour identification capabilities~\cite{Akindinov:2019flp}.} of P2O experiment using a 90 kW proton beam 
with a runtime of 3 yrs. in $\nu$ and 3 yrs. in $\bar{\nu}$ mode, - corresponding to a total POT of $4.8\times 10^{20}$. 
The baseline mostly passes through the upper mantle of the earth with an average 
density of $3.4$ g/cc and the deepest point along the beam being 134 km~\cite{p2o_brunner_talk_2021}.
The fluxes, detector response parameters, the detection efficiencies, signal and background systematics \etc, corresponding to our nominal P2O configuration were taken from \cite{Akindinov:2019flp, Adrian-Martinez:2016fdl}. 

%The majority of neutrino events observed by ORCA will be from electron and muon neutrino and antineutrino charge-current (CC) interactions, while tau neutrinos and neutral current (NC) interactions constitute minor backgrounds. 
%Currently, 6 ORCA lines are live and taking data since a year.

To estimate the sensitivity of LBL experiments to probe the LIV parameters, we carry out a $\chisq$ analysis 
using GLoBES. 
The analytical\footnote{This is the {\it{Poissonian}} definition of $\chisq$, which in the limit of large sample size, reduces to the Gaussian form.} form of the $\chisq$ can be expressed as,
 \begin{align}
\label{eq:chisq}
\Delta \chi^{2}(p^{\text{true}}) = \underset{p^{\text{test}}, \eta}{\text{Min}} \Bigg[&2\sum_{k}^{\text{mode}}\sum_{j}^{\text{channel}}\sum_{i}^{\text{bin}}\Bigg\{
N_{ijk}^{\text{test}}(p^{\text{test}}; \eta) - N_{ijk}^{\text{true}}(p^{\text{true}})
+ N_{ijk}^{\text{true}}(p^{\text{true}}) \ln\frac{N_{ijk}^{\text{true}}(p^{\text{true}})}{N_{ijk}^{\text{test}}(p^{\text{test}}; \eta)} \Bigg\}  \nonumber \\
&+ \sum_{l}\frac{(p^{\text{true}}_{l}-p^{\text{test}}_{l})^{2}}{\sigma_{p_{l}}^{2}}
+ \sum_{m}\frac{\eta_{m}^{2}}{\sigma_{\eta_{m}}^{2}}\Bigg].
\end{align}

 $N^{\text{true}}$ corresponds to the simulated set of event spectra corresponding to \textit{true} set of oscillation parameters $p^{\text{true}}$, 
 where only standard scenario is assumed with all the LIV parameters $a_{\alpha\beta}$ ($\alpha, \beta = e, \mu, \tau$) kept fixed to zero and all the standard oscillation parameters are kept fixed to their bestfit values.
 $N^{\text{test}}$ denotes the events simulated in presence of LIV, 
 where the LIV parameters, as well as some of the less well-measured standard oscillation parameters are allowed to vary. 
 %%%%%%%%%%%%%%%%%%%%%%%%
 \begin{table}[th]
\centering
\scalebox{0.8}{
\begin{tabular}{|c| c | c | c | c |}
\hline
%&&&\\
Fixed/Varied & Parameter & Bestfit/true value & Variation range & $1\sigma$ uncertainty (prior)  \\
&&($p^\text{true}$)&$(3\sigma$ interval) ($p^\text{test}$)& ($\sigma_{p_{l}}$)\\
\hline
%&&&\\
Fixed & $\theta_{12}$ [Deg.]             & 34.3                    &  $[31.4, 37.4]$   &  - \\
  & $\theta_{13}$  [Deg.]    & 8.53              &  $[8.16, 8.94]$   &  - \\
 & $\sdm$ [$10^{-5}\text{ eV}^2$]  & $7.5  $  &  $[6.94, 8.14]$  &  - \\
\hline
Varied & $\theta_{23}$ [Deg.]        & 48.8                     &  $[41.63, 51.32]$    &  3.5\% \\
  & $\ldm$  [$10^{-3}\text{ eV}^2$] & $2.55$   &  $[2.46, 2.65] \cup [-2.55, -2.37]$ &  2.4\% \\
 & $\delta_{13}$  [Deg.]   & $-122.4$   & $[-180, 0]  \cup [144, 180]$ &  - \\
& $\aee [10^{-23} \text{ GeV}]$ & 0 & $[-40, 40]$ & -\\
& $\amm [10^{-23} \text{ GeV}]$ & 0 & $[-10, 10]$ & -\\
& $|\aem| [10^{-23} \text{ GeV}]$ & 0 & $[0, 5]$ & -\\
& $|\aet| [10^{-23} \text{ GeV}]$ & 0 & $[0, 5]$ & -\\
& $|\amt| [10^{-23} \text{ GeV}]$ & 0 & $[0, 5]$ & -\\
& $\phem$ [Deg.] & 0 & $[-180, 180]$ & -\\
& $\phet$ [Deg.] & 0 & $[-180, 180]$ & -\\
& $\phmt$ [Deg.] & 0 & $[-180, 180]$ & -\\
\hline
\end{tabular}}
\caption{\label{tab:parameters}
The values of standard and LIV parameters used in our study. 
 The first column indicates whether the parameters were kept fixed or varied around their true values. 
 The third column shows the {\it{true}} values used (taken from the globalfit analysis in \cite{deSalas:2020pgw}), while 
 the next column shows the range of variation (taken to be the current $3\sigma$ interval). 
 The rightmost column shows the prior uncertainties used while varying the corresponding parameters in the analysis.   
If the $3\sigma$ upper and lower limit of a parameter is $x_{u}$ and $x_{l}$ respectively, the $1\sigma$  uncertainty is $(x_{u}-x_{l})/3(x_{u}+x_{l})\%$~\cite{Abi:2020evt}.
}
\end{table}
%%%%%%%%%%%%%%%%%%%%%%%%
 The total set of standard and LIV parameters that generate $N^{\text{test}}$ are denoted by $p^{\text{test}}$.
Table \ref{tab:parameters} summarizes the values of the standard and LIV oscillation parameters used in our analysis. 
Note that in generating $N^{\text{test}}$ we have kept the three well-measured standard parameters $\ta, \tb, \sdm$ fixed to their bestfit values.  
We have checked that varying these three parameters in the fit produces negligible changes to the result.
We varied the other three less well-measured standard parameters $\tc, \ldm, \dcp$, as well as the LIV parameters $|a_{\alpha\beta}|, \varphi_{\alpha\beta}$ ($\alpha, \beta = e, \mu, \tau$).
Throughout the analysis we assume the {\it{true}} mass hierarchy to be normal and vary the {\it{test}} value of $\ldm$ over both the normal and inverted hierarchy.
 The sums over the three indices $i, j, k$ signify the summations over the energy bins, 
 the oscillation channels ($\nu_{e}$ appearance and $\nu_{\mu}$ disappearance), and the 
 running modes (neutrino and antineutrino modes) respectively.  
For DUNE we take a total of 71 energy bins in the range of $0-20$ GeV, - with $64$ bins 
with uniform widths of 
$0.125$ GeV in the energy range of 0 to 8 GeV and $7$ bins
with varying widths beyond $8$ GeV~\cite{Alion:2016uaj}. 
For P2O, we take 40 uniform bins up to 12 GeV. 
Thus the first term ($N^{\text{test}} - N^{\text{true}}$) inside the curly braces accounts for the algebraic difference between the two sets of data, whereas the log-term gives a kind of fractional difference between them.
The entire expression in the curly brackets with summations over $i, j, k$ consists of the statistical part of the $\chisq$.
%%%%%%%%%%%%%%%%%%%%%%%%
 \begin{table}[h]
\centering
\scalebox{0.9}{
\begin{tabular}{| c | c | c |}
\hline
%&&&\\
Systematics & Uncertainty ($\sigma_{\eta}$) & Uncertainty ($\sigma_{\eta}$)   \\
/Nuisance parameters ($\eta$) & (DUNE) & (P2O) \\
%&&&\\
\hline
%&&&\\
$\nu_{e}$ signal normalization & $2\%$ & $5\%$ \\
$\bar{\nu}_{e}$ signal normalization & $2\%$ & $5\%$ \\
$\nu_{\mu}$ signal normalization & $5\%$ & $5\%$ \\
$\bar{\nu}_{\mu}$ signal normalization & $5\%$ & $10\%$ \\
%&&&\\
\hline
$\nu_{e}$ background normalization & $5\%$ & $10\%$ \\
$\bar{\nu}_{e}$ background normalization & $5\%$ & $10\%$ \\
$\nu_{\mu}$ background normalization & $5\%$ & $10\%$ \\
$\bar{\nu}_{\mu}$ background normalization & $5\%$ & $10\%$ \\
Neutral current background normalization & $10\%$ & $10\%$ \\
$\nu_{\tau}$ background normalization & $20\%$ & $20\%$ \\
$\bar{\nu}_{\tau}$ background normalization & $20\%$ & $20\%$ \\
\hline
Density & $10\%$ & $10\%$\\
\hline
\end{tabular}}
\caption{\label{tab:nuisance}
The signal/ background systematics and the density uncertainties used in our analysis for both DUNE and P2O configurations.
}
\end{table}
%%%%%%%%%%%%%%%%%%%%%%%%%%%%

Uncertainties in the prior measurement of the $l^{th}$ oscillation parameter are given by the parameters $\sigma_{pl}$. 
As indicated in Table \ref{tab:parameters}, for the variation of $\tc$ and $\ldm$, we have used prior uncertainties of  $3.5\%$ and $2.4\%$ respectively\footnote{{Due to the degenerate solutions existing in two octant of $\tc$, we have also checked some of our subsequent results by increasing the prior uncertainty of $\tc$ to values higher than $3.5\%$, such that the {\it{test}} $\tc$-values span both the octants. We found that as long as the prior is not too high ($\lesssim 15\%$), the results do not change significantly.}}. 
For $\ldm$ we have varied over the sign also to take care of the possible fake solutions in the opposite mass hierarchy.
We have allowed the rest of the parameters $\dcp, \aee, \amm, |\aem|, |\aet|, |\amt|, \phem, \phet, \phmt$ to vary in an unrestricted manner without any prior uncertainties.
$\eta_{m}$ is the nuisance parameter/systematics and $\sigma_{\eta_{m}}$ is the corresponding uncertainty which arises from the detector properties. 
Table \ref{tab:nuisance} summarizes the overall normalization uncertainties of these systematic parameters for various signals and backgrounds used in our analysis. 
We assume the various signal and background systematic parameters are distributed in a Gaussian way with mean value 0 and standard deviation $\sigma_{\eta_{m}}$, indicated in the second and third columns of Table \ref{tab:nuisance} for DUNE and P2O respectively.
This way of treating the systematics in the $\chisq$ calculation is known as the {\it{method of pulls}}~\cite{Huber:2002mx,Fogli:2002pt,GonzalezGarcia:2004wg,Gandhi:2007td}.
The background normalization uncertainties include correlations among various sources of backgrounds (contamination of $\nu_{e}/\bar{\nu}_{e}$ in the incident beam, flavour misidentification, neutral current, and $\nu_{\tau}$). 
We further include a $10\%$ prior uncertainty on both the baseline densities of DUNE ($2.95$ g/cc) and P2O ($3.4$ g/cc).
%The term summed over $l$ takes care of the precision measurements in the oscillation parameters and the last term (summed over $m$) is the systematics part. 
The final estimate of the (minimum) $\chisq$ is obtained after varying the relevant oscillation parameters  (as mentioned earlier and summarized in Table \ref{tab:parameters} and the systematic parameters along with the densities (see Table \ref{tab:nuisance}), and reporting the minimum value of the $\chisq$. 
The procedure is known as the marginalization of the relevant oscillation parameters so that the final result gives a conservative estimate of $\chisq$.
The $\chisq$ thus estimated is the frequentist method 
of hypotheses testing~\cite{Fogli:2002pt, Qian:2012zn}.

%==============================================
 \section{Correlations among the LIV parameters}                         %Section V
 \label{sec:LIV_corr}
 %================================================
  %%%%%%%%%%%%%%%%%%%%%
  \begin{figure}[thb]
 \centering
 \includegraphics[width=0.9\textwidth, height=1\textwidth]{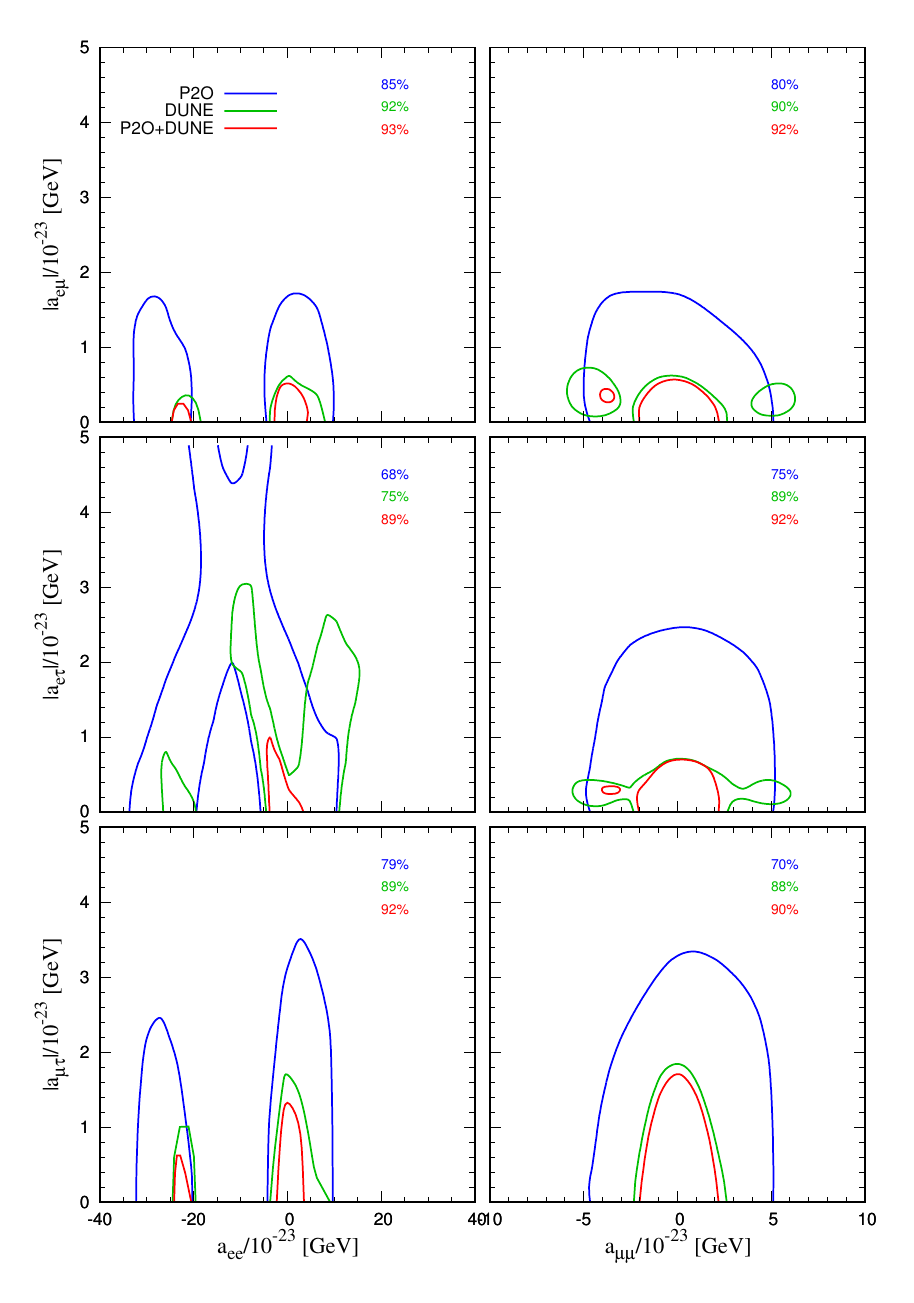}
 \caption{\footnotesize{This shows the exclusion regions in the parameter space consisting of one diagonal (along the horizontal axis)
 and one off-diagonal LIV parameter (vertical axis) for P2O only (blue contours), DUNE only (green contours), and P2O combined with DUNE (red contours). 
 The results are shown at the confidence level (C.L.) of $95\%$.  
 The triplet of numbers ($\%$) in each panel indicates the area lying (excluded) outside the $95\%$ C.L. contours, expressed as a percentage of the total area of the parameter space considered. 
 The numbers are shown for the three cases, - blue for P2O only, green for DUNE only, and red for the combined case of (P2O + DUNE), and thus they offer a measure of the exclusion capabilities of each experimental configuration for each relevant parameter space.}}
  \label{fig:chisq_ab_aa}
 \end{figure}
 %%%%%%%%%%%%%%%%%%%%%%
 In Fig.\ \ref{fig:chisq_ab_aa}, we show the $95\%$  confidence level (C.L.) regions in 
 the parameters space spanned by one off-diagonal LIV parameter $|a_{\alpha\beta}|$ ($|\aem|, |\aet|$ or $|\amt|$) and one diagonal LIV parameter $a_{{\alpha}^{\prime}{\beta}^{\prime}}$ ($\aee$ or $\amm$).  
Thus we assume the presence of two LIV parameters at a time in the fit. The three standard parameters as well as the relevant LIV phases are then varied (see Table \ref{tab:parameters}) and to obtain the minimum $\chisq$ (\ie, we marginalize over the three standard parameters and the LIV CP phases.). 
 For instance, for the analysis in the parameter space of $\aee-|\aem|$, we vary $\tc, \ldm$ (sign and magnitude) with priors $3.5\%$ and $2.4\%$ respectively and $\dcp, \phem$ without priors in an unrestricted manner and obtain the minimum $\chisq$ as a function of $\aee$ and $|\aem|$. We repeat the procedure for many sets of $(\aee, |\aem|)$ to plot the iso-$\chisq$ contours at a C.L. of $95\%$ (which corresponds to a $\chisq$ value of 5.99 for 2 d.o.f.).
 The blue contours show the sensitivity reach of P2O alone while the red ones illustrate the results of 
 combining the projected data of  P2O and DUNE (which we write as (P2O + DUNE) hereafter). 
 For completeness we have also shown the analysis with DUNE only case, although 
 that is not the main focus of our work. We refer the interested readers to \cite{Barenboim:2018ctx} for a more comprehensive analysis of the capabilities of DUNE to probe LIV parameter space\footnote{The analysis with DUNE-only case in the present work is qualitatively consistent with \cite{Barenboim:2018ctx} with some minor differences due to different choices of the set of values of the oscillation parameters, different ranges of marginalization, different minimization techniques \etc}.
For each of these three experimental configurations, namely P2O, DUNE, and (P2O+DUNE), in each panel, we estimate the regions excluded at $95\%$ C.L. contours (\ie, the area outside the contours with blue, green, and red boundaries respectively), and express that as a percentage of the total area of 
the parameter space is shown. 
The three numbers thus give us a quantitative measure of the improvement of the combination 
(P2O + DUNE) over P2O only or DUNE only in excluding the relevant parameter space at $95\%$ C.L.\footnote{Similar estimates were used in reference \cite{Masud:2018pig} in order to quantify the improvement 
of one experimental configuration over another in the context of non-standard neutrino interaction.}
The improvement is remarkable in almost all cases, covering more than $90\%$ of the parameter 
space considered. 
In presence of $\aee$, we observe the additional/fake degenerate region around $\aee \simeq -22 \times 10^{-23}$ GeV, which arises due to marginalization over the opposite mass hierarchy. 
Note that the location of this fake solution is approximately opposite in sign to the degeneracies in the corresponding probability 
heatplots (Figs.\ \ref{fig:heatplot_dcp} and \ref{fig:heatplot_th23}: third column, top row), 
where the additional degeneracies were found around $\aee \simeq 22 \times 10^{-23}$.
It can be qualitatively understood as follows. 
Without considering flux and cross-sections for simplicity, the dominant statistical contribution to the sensitivity in the LIV scenario ({\textit{test}} scenario) involving the parameter $\aee$ and another parameter, say c,  
roughly follows the corresponding probability deviation from the {\textit{true}} standard case 
(in the similar spirit as discussion in Sec.\ \ref{sec:prob_LIV}): 
\begin{equation}\label{eq:chisq_aee}
\chisq (\aee, c) \sim \dpme (\aee) + \dpme (c) + (\text{other terms}),
\end{equation}
where the other terms contain contributions from the $\numu$ disappearance channel, 
antineutrinos, priors and systematics, - which we have neglected in order to have a simple 
qualitative understanding.
%%%%%%%%%%%%%%%%%%%%%
 \begin{figure}[h]
 \centering
 \includegraphics[scale=1.2]{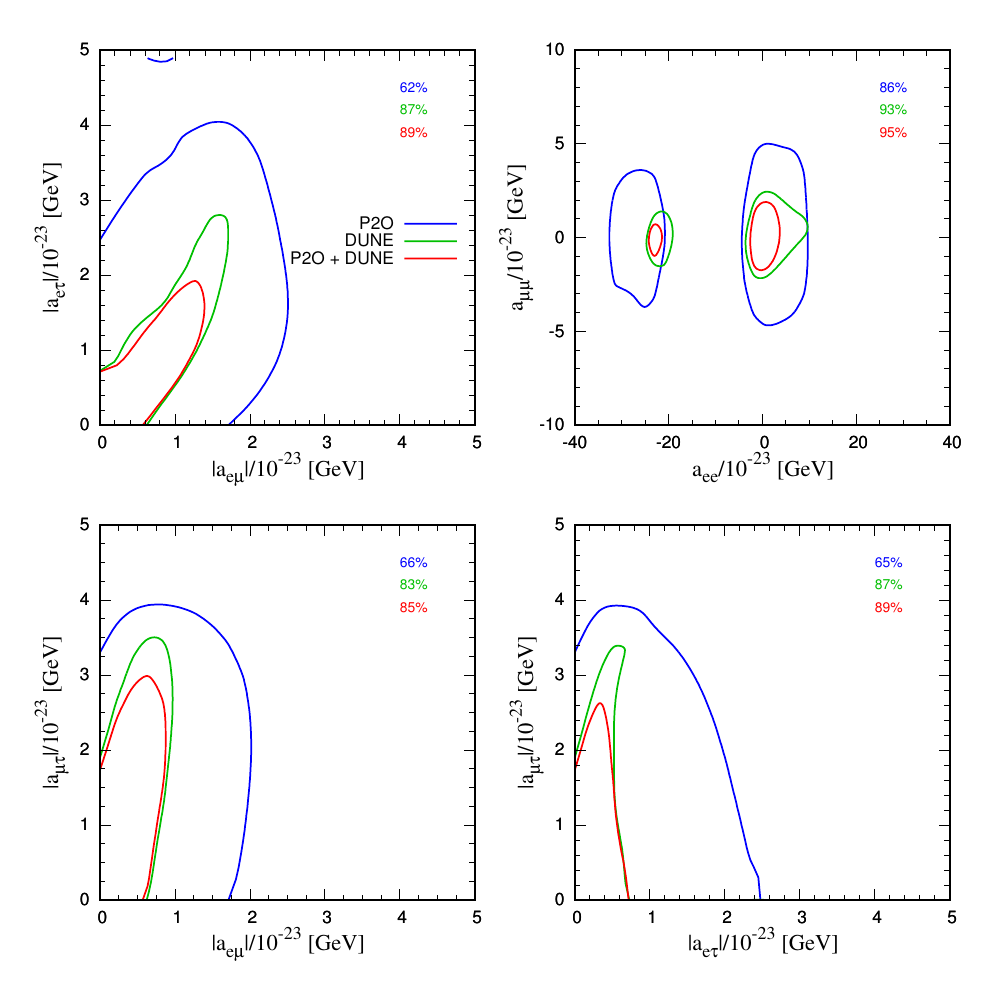}
 \caption{\footnotesize{Similar to Fig.\ \ref{fig:chisq_ab_aa} but shows the exclusion regions 
 in the parameter spaces with both off-diagonal LIV parameters (top-left, bottom-left and bottom-right panels) and both diagonal LIV parameters (top right panel).}}
  \label{fig:chisq_ab_ab}
 \end{figure}
%%%%%%%%%%%%%%%%%%%%%%%
Using our previous discussion concerning Eqs.\ \ref{eq:dpme_aee} and \ref{eq:zero_dpme_aee}, we can write,
 \begin{align}\label{eq:chisq_aee_1}
&\chisq (\aee, c) 
\sim \dpme (\aee) \nonumber \\
&\sim 
\Bigg[
 \underbrace{
\frac{\sin\big[1-\hat{A}(1+\hat{a}_{ee})\big]\Delta}
{1-\hat{A}(1+\hat{a}_{ee})}
- \frac{\sin\big[1-\hat{A}\big]\Delta}
{1-\hat{A}} 
}_{\text{I}_{-}}
\Bigg]
\times \Bigg[
\underbrace{
\frac{\sin\big[1-\hat{A}(1+\hat{a}_{ee})\big]\Delta}
{1-\hat{A}(1+\hat{a}_{ee})}
+\frac{\sin\big[1-\hat{A}\big]\Delta}
{1-\hat{A}} 
}_{\text{I}_{+}}
\Bigg]. 
\end{align}
Within the same mass hierarchy for the true and test scenario, the minimum for $\chisq (\aee, c)$ is obtained at the true 
solution $\aee \simeq 0$, making $\text{I}_{-}$ vanish. 
But while marginalizing over the opposite mass hierarchy in the test scenario, $\hat{A}$ and $\Delta$ changes sign in the term $\sin[1-\hat{A}(1+\hat{a}_{ee})]\Delta
/[1-\hat{A}(1+\hat{a}_{ee})]$ containing $\aee$, and we have, 
\begin{align}\label{eq:chisq_aee_2}
\chisq (\aee, c) \sim 
\Bigg[
 \underbrace{
\frac{\sin\big[1+\hat{A}(1+\hat{a}_{ee})\big]\Delta}
{1+\hat{A}(1+\hat{a}_{ee})}
+ \frac{\sin\big[1-\hat{A}\big]\Delta}
{1-\hat{A}} 
}_{\text{I}_{-}}
\Bigg]
\times \Bigg[
\underbrace{
\frac{\sin\big[1+\hat{A}(1+\hat{a}_{ee})\big]\Delta}
{1+\hat{A}(1+\hat{a}_{ee})}
-
\frac{\sin\big[1-\hat{A}\big]\Delta}
{1-\hat{A}} 
}_{\text{I}_{+}}
\Bigg].
\end{align}
Because of the relative changes of signs, now $\text{I}_{-}$ cannot vanish and the minimum solution is obtained when $\text{I}_{+}$ goes to zero instead. 
That is obtained when $\hat{a}_{ee} = -2$ and thus $\aee \simeq -22 \times 10^{-23}$ GeV.  
Such a degeneracy can also be observed for the DUNE only case (green contours),   which is consistent with previous analyses with LIV in case of DUNE~\cite{Barenboim:2018ctx}.
%On closer inspection at the $|\aet|-\aee$ plane (second row, first column of Fig.\ \ref{fig:chisq_ab_aa}), we note an interesting correlation. 
%There are two sets of degenerate regions (around $\aee \simeq 0$ and $\aee \simeq -22 \times 10^{-23}$ GeV) each containing further couple of branches. 
% Such a degeneracy appears due to the fact that $\aee$ and $|\aet|$ changes the probability 
% in the opposite direction (see Fig.\ \ref{fig:prob}). 
 Although a combination with DUNE significantly constrains this additional degeneracy at $95\%$ C.L., it still does not 
 go away completely (we have checked that it still remains at $99\%$ C.L.). 
 For the parameter $\amm$ we see the contours are roughly symmetric around the true solution $\amm = 0$. 
 If $\amt$ is present, a combination with DUNE can probe more than $92\%$ of the entire parameter space considered at a C.L. of $95\%$. 
 This sensitivity to $|\amt|$ mainly comes from the $\numu$ disappearance channel.

 Fig.\ \ref{fig:chisq_ab_ab} shows the $\chisq$ correlation among the off-diagonal 
 LIV parameters themselves ($|\aem|, |\aet|, |\amt|$) and also between the two diagonal 
 parameters $\aee$ and $\amm$, for P2O, DUNE and (P2O+DUNE). 
 The improvement by the combined analysis is especially prominent for the most 
 impactful parameter space $\aem-\aet$ (top left panel of Fig.\ \ref{fig:chisq_ab_ab}). 
 At a C.L. of $95\%$, (P2O+DUNE) combination can exclude $89\%$ of the parameter 
 ranges considered, compared to $62\%$ by P2O alone.  
   
 %==============================================
 \section{Degeneracies with the standard oscillation parameters}                         %Section VI
 \label{sec:LIV_std_deg}
 %================================================
%%%%%%%%%%%%%%%%
 \begin{figure}[h]
 \centering
 \includegraphics[scale=1.2]{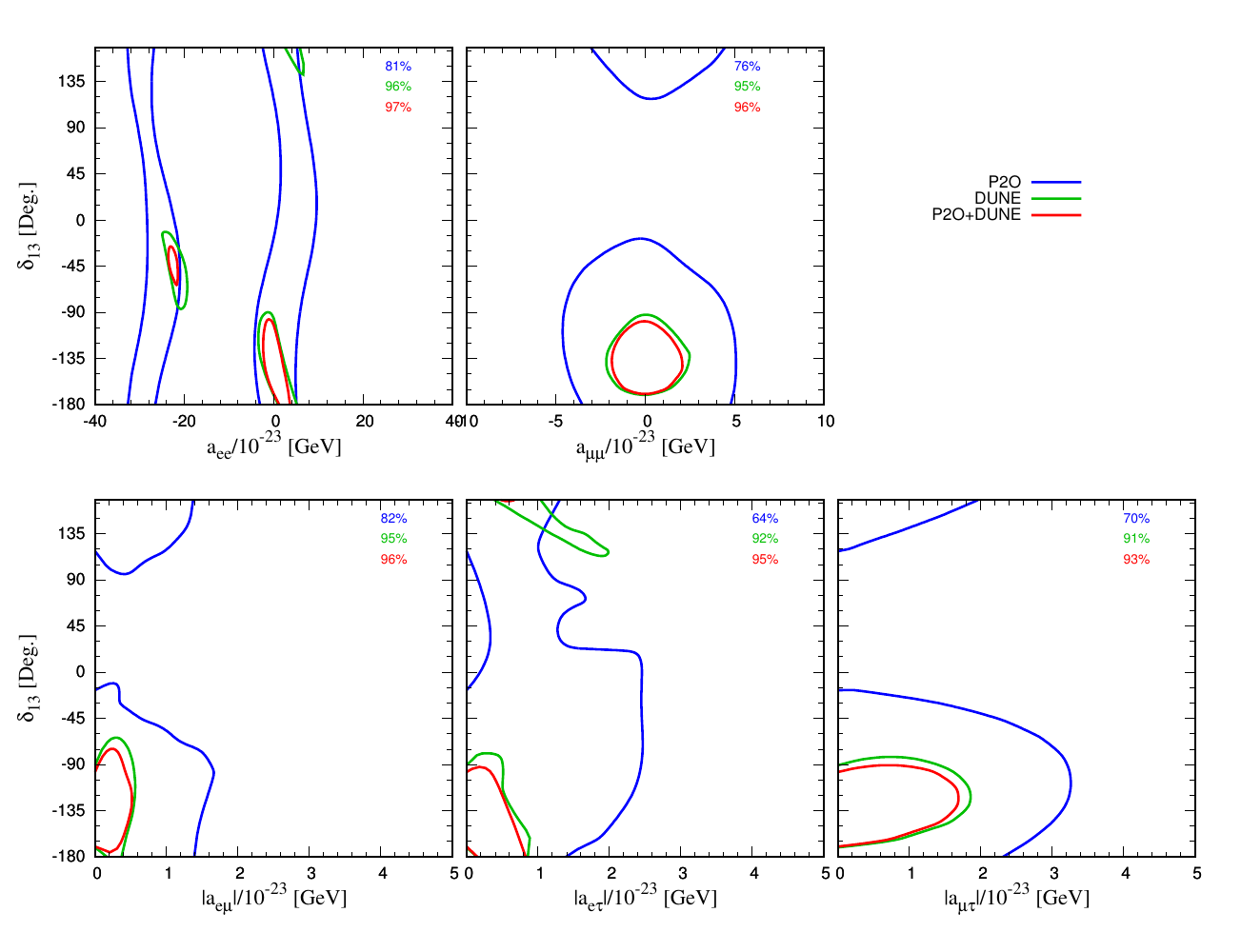}
 \caption{\footnotesize{Exclusion regions at $95\%$ C.L.,  
 showing the correlations of the LIV parameters with the CP phase $\delta_{13}$ for P2O (blue contours), DUNE (green) and P2O+DUNE (red contours). 
Similar to Fig.\ \ref{fig:chisq_ab_aa}, the triplet of numbers ($\%$) in each panel indicates the area of parameter space excluded at $95\%$ C.L., expressed as a percentage of the total area of the parameter space considered.}}
  \label{fig:chisq_corr_dcp}
 \end{figure}
 %%%%%%%%%%%%%%%
In Fig.\ \ref{fig:chisq_corr_dcp} (and Fig.\ \ref{fig:chisq_corr_th23}), we demonstrate how efficiently
 the projected data from P2O, DUNE and the combined case of (P2O+DUNE) can reconstruct the standard CP phase $\dcp$ and mixing angle $\tc$, in correlation with the LIV parameters present. 
Here we assume the presence of one LIV parameter at a time and marginalize over the relevant LIV phase, as well as 
 over the standard parameter not shown along the axes (see Table \ref{tab:parameters} for the ranges and priors.). 
 For instance, for the analysis in the $|\aem|-\dcp$ plane, the minimum $\chisq$ is obtained after varying $\ldm$ (both magnitude  and sign) and $\tc$ with priors of $2.4\%$ and $3.5\%$ respectively and $\phem$ without prior in an unrestricted manner. Similarly, for the $|\aem|-\tc$ plane, the marginalization is carried over $\dcp$ and $\phem$ without priors in an unrestricted way and over $\ldm$ with a $2.4\%$ prior.
% The true value of $\tc$ was taken as $48.8^{\circ}$ and in the test dataset the marginalization was done on $\tc, \dcp, \ldm$ (both magnitude and the sign), as well as the LIV CP phases wherever present. 
 At the C.L. of $95\%$, the presence of any LIV parameter at P2O can give rise to 
 allowed regions covering a large $\dcp$-space (entire $\dcp$-space for $\aee$ and $\aet$.). 
 But the combination (P2O + DUNE) significantly shrinks the allowed regions to lie around 
 the true solution of $\dcp = -122.4^{\circ}$. 
 Similar observation holds in Fig.\ \ref{fig:chisq_corr_th23} for the parameter space containing $\tc$ also. 
Concerning the combined analysis of (P2O+DUNE) (\ie, the red contours in Fig.\ \ref{fig:chisq_corr_th23}), although the maximal mixing ($\tc = 45^{\circ}$) got excluded in case of all the LIV parameters, in case of $\aem$ and $\aet$ (bottom row, first and second columns of Fig.\ \ref{fig:chisq_corr_th23}) we note that the allowed regions still appear in the opposite (lower) octant. 
 We refer the readers to \cite{KumarAgarwalla:2019gdj} for a more in-depth discussion 
 regarding the impacts of $\aem$ and $\aet$ on $\tc$-octant. 
 It is clear that for P2O alone, the exclusion region in presence of $\aem$ is greater than in presence of $\aet$ 
 ($67\%$ versus $54\%$ of the total parameter space considered, at $95\%$ C.L.). 
 This can be connected to the higher impact of $\aem$ in the probability deviation $|\dpme|$ in Fig.\ \ref{fig:heatplot_th23} (top row, first and second column) and related discussions in Sec.\ \ref{sec:prob_LIV}. 
 $\aee$ generates the additional degeneracy around $-22 \times 10^{-23}$ GeV as usual.
% The LIV parameters are varied along the x-axis and the CP phase is varied along the y-axis encompassing the entire range $[-\pi, \pi]$.
% At $3\sigma$ confidence level P2O simulated data fails to give any bound on the values of CP phase but has tighter bounds in case of $a_{ee}$ and $a_{e\mu}$.
% The $2\sigma$ contour for P2O simulated data tightens the bounds for the LIV parameters, also, shows some exclusion regions for $\delta_{CP}$ in correlation with $a_{\mu \beta} (\beta = e, \mu, \tau)$.
% Adding DUNE data changes the scenario drastically.
% The allowed regions shrink considerably around the true values in the parameter space.
% We still have the degeneracy along $a_{ee}$.
% The contours in the bottom panel (with the off-diagonal terms of the LIV parameters) are not closed because we have taken the  modulus($|a_{\alpha\beta}|$) of these parameters while marginalising over the phase factors ($\phi_{\alpha \beta}$) which take care of the sign of these parameters.
%%%%%%%%%%%%%%%%%%%%% 
  \begin{figure}[h]
 \centering
 \includegraphics[scale=1.2]{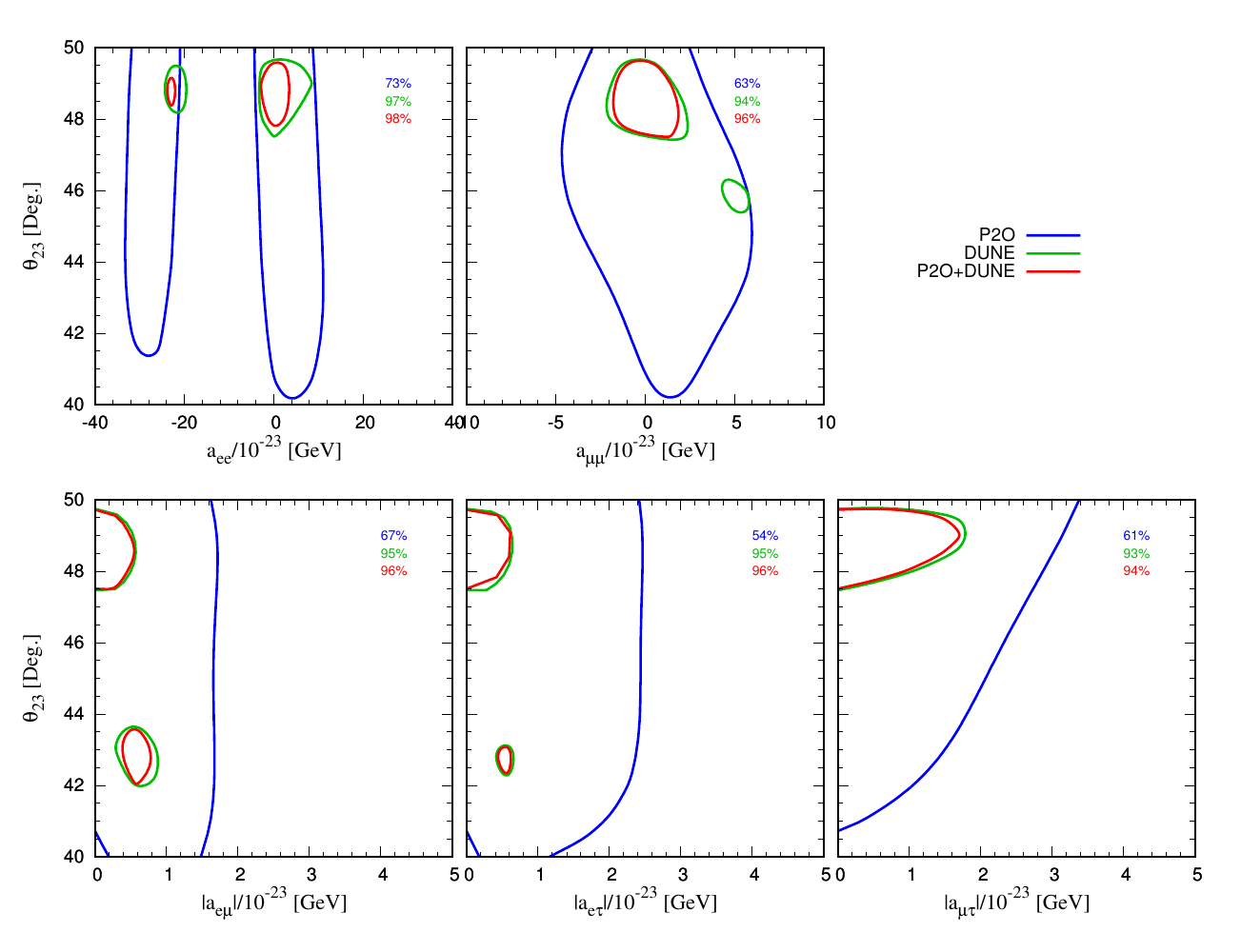}
 \caption{\footnotesize{Similar to Fig.\ \ref{fig:chisq_corr_dcp} but showing the correlation of the LIV parameters with $\tc$.}}
  \label{fig:chisq_corr_th23}
 \end{figure}
 %%%%%%%%%%%%%%%%%%%%%%%
%Here in Fig.\ \ref{fig:chisq_corr_th23} in spite of choosing a higher octant \textit{true} value for $\theta_{23}$ we get a degeneracy in the lower octant region in case of $a_{e \mu}$ and $a_{e \tau}$ reconstruction in spite of taking $\theta_{23,\ true} = 48.8^{\circ}$
%Even then, the value for maximal mixing (\ie $\theta_{23} = 45^{\circ}$) lies in the exclusion region.
%But in case of $a_{ee}$, $a_{\mu \mu}$ and $a_{\mu \tau}$ the degeneracy in $\theta_{23}$is lifted when we combine DUNE and P2O simulated data.
%The allowed region discards the lower octant values.
%
%We already know the bounds on the standard oscillation parameters as shown in Tab [2.1].
%If the phenomenon of LIV is true in nature, the P2O simulated data along with DUNE can lift the fake octant degeneracy in $\theta_{23}$ in correlation to the LIV parameters. 

%==============================================
 \section{Bounds on the LIV parameters}\label{sec:bound}                                                                                                             %Section VII
 %================================================
%%%%%%%%%%%%%%%%%%%%%%
  \begin{figure}[h]
 \centering
 \includegraphics[scale=0.55]{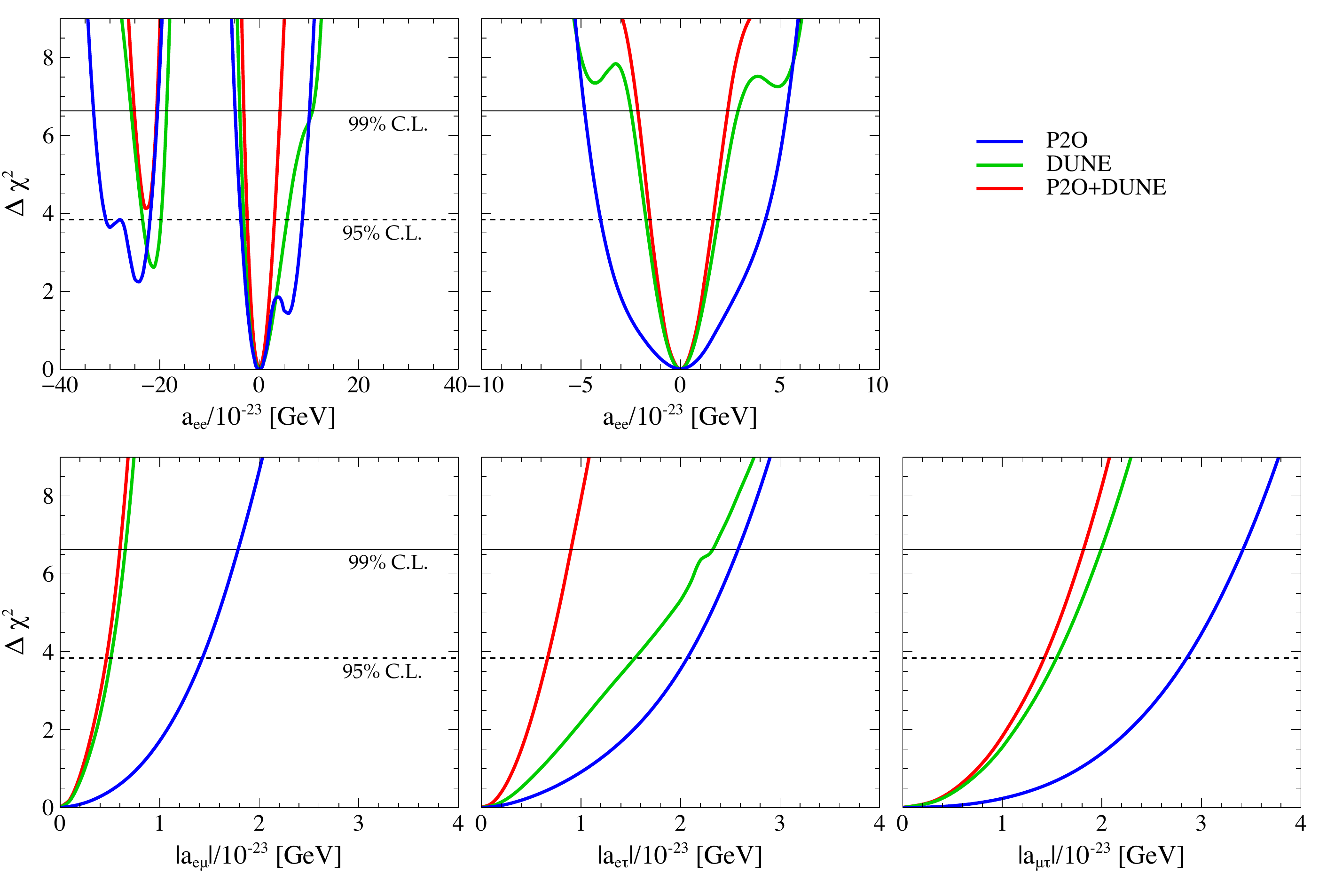}
 \caption{\footnotesize{Expected sensitivity of P2O (blue) and the combined (P2O+DUNE) (red) analysis  to the LIV parameters. The black dotted (solid) line indicates the $95\%$ ($99\%$) C.L. at 1 degree of freedom. Each panel corresponds to individual LIV parameter.}}
  \label{fig:chisq_1param}
 \end{figure}
%%%%%%%%%%%%%%%%%%%%%%
%%%%%%%%%%%%%%%%%%%
 \begin{table}[h]
\centering
\scalebox{0.9}{
\begin{tabular}{| c | c | c | c |}
\hline
&&&\\
Parameter & Bounds from DUNE  & Bounds from P2O & Bounds from (P2O+DUNE) \\ 
& [$10^{-23}$ GeV]  & [$10^{-23}$ GeV] & [$10^{-23}$ GeV] \\
\hline
&&&\\
$a_{ee}$            
 &  $[-24 <  a_{ee} < -20]$       
    &  $[-30.8 <  a_{ee} < -21.9]$  
    & $-2.6  <  a_{ee} < 3.3 $ \\
    & $\cup\ [-3.2 <  a_{ee} < 5.6]$ & $\cup\ [-3.9 <  a_{ee} < 8.6] $ &\\
$a_{\mu \mu}$    
&  $-1.9 <  a_{\mu \mu} < 2.0 $            
 &  $-4.0  <  a_{\mu \mu} < 4.3 $ 
 &  $-1.6  <  a_{\mu \mu} < 1.6 $\\
$|a_{e \mu}| $   & $0.6 $             &  $1.6 $ &  $0.4 $ \\
$|a_{e \tau}| $  & $1.3 $             &  $2.1 $ &  $0.7 $ \\
$|a_{\mu \tau}| $   & $1.5 $              &  $2.9$ &  $1.3$  \\
\hline
\end{tabular}}
\caption{\label{tab:LIV_parameters} Bounds on the LIV parameters as obtained from the
simulations of LBL data: (DUNE, P2O and (P2O+DUNE) in the 2nd, third, fourth column respectively) at 95\% C.L.
 }
\end{table}
%%%%%%%%%%%%%%%%%%%%%%%%%%%
 Fig.\ \ref{fig:chisq_1param} shows the one dimensional projection (after marginalising away all other parameters including the CP phases, $\tc$, $\ldm$) of $\chisq$ as a function of each LIV parameter individually. 
 The results are illustrated for P2O alone (blue), DUNE alone (green) and for the combined analysis of (P2O + DUNE) (red). 
 The $\chisq$ values corresponding to $95\%$ and $99\%$ C.L.s are marked with 
 horizontal black lines. 
 The significant increase in the steepness of the red sensitivity curves is indicative of the crucial impact of the combined analysis in constraining the LIV parameters.  
 For $\aee$, we note the lifting of the troublesome degeneracy at $\aee \simeq -22 \times 10^{-23}$ GeV by the combined analysis above $95\%$ C.L. This was not possible with 
 the analysis done with DUNE alone (see also ~\cite{Barenboim:2018ctx}) or with P2O alone.
 Table \ref{tab:LIV_parameters} shows our final result: the constraints obtained on the five 
 LIV parameters at $95\%$ C.L. with the combined (P2O+DUNE) analysis and compares the numbers 
 obtained from only DUNE or only P2O. 
 We note that the constraints on the diagonal parameters can be tightened significantly 
 with the combined analysis. 
 This is especially noticeable for $\aee$ since the fake solution can be ruled out at $95\%$ C.L. as mentioned before. 
 For $|\aem|$, $|\aet|$ and $|\amt|$ also the bounds improve moderately. 
 Some comments are in order regarding the bounds on LIV parameters achieved by existing atmospheric neutrino data. Atmospheric neutrino experiments are sensitive to a much wider range of baseline and energy, and hence can obtain strong constraints on LIV parameters. For instance, the SK data have put the following  bounds on the LIV parameters at $95\%$ C.L.~\cite{Super-Kamiokande:2014exs}, 
 \begin{align}
 &|\amm| \lesssim 1.9 \times 10^{-23} \text{ GeV}, \nonumber \\
 &\text{Re}(\aem), \text{Im}(\aem) \lesssim 1.8 \times 10^{-23} \text{ GeV}; \nonumber \\
 &\text{Re}(\aet) \lesssim 4.1 \times 10^{-23} \text{ GeV}; 
 \text{Im}(\aet) \lesssim 2.8 \times 10^{-23} \text{ GeV}; \nonumber 
 \end{align}
 which roughly translate to,
 \begin{align}
& |\aem| \lesssim 3.2 \times 10^{-23} \text{ GeV}; \nonumber \\
 &|\aet| \lesssim 5 \times 10^{-23} \text{ GeV}. \nonumber
 \end{align}
 Analysis of high energy astrophysical and atmospheric neutrino at IceCube have put following stronger constraints, at an even higher statistical significance of $99\%$ C.L.~\cite{IceCube:2017qyp}:
 \begin{equation}
 |\text{Re}(\amt)|, |\text{Im}(\amt)| \lesssim 0.29 \times 10^{-23} \text{ GeV} 
 \implies |\amt| \lesssim 0.41 \times 10^{-23} \text{ GeV}. \nonumber
 \end{equation}
 The next generation IceCube-Gen2~\cite{IceCube:2014gqr} is expected to reach much tighter bounds on LIV parameter space. 
 On the other hand if LBL experiments such as DUNE is also used to collect 
 and analyse atmospheric neutrino data (in addition to beam neutrinos), it becomes sensitive to a much wider range of baseline and energy, and the constraints on LIV 
 parameters can then improve by several orders of magnitude~\cite{Abi:2020evt}. We would also like to mention that in comparison to LBL data, high energy astrophysical and atmospheric neutrino experiments can become more sensitive to higher order LIV parameters (which are energy-dependent) which we have not considered in the present analysis. 
 As more neutrino data become available in near future, it will be possible to strategically combine LBL and atmospheric neutrino data and search for the presence of LIV with an unprecedented sensitivity reach. 
 However, in the present analysis we have focused on the capability of LBL experiments only to probe the LIV parameters. 
 A full combined analysis of both LBL and atmospheric data (simulated/real data) is beyond the scope of the current work, and we leave it as a future project.
 
%==============================================
 \section{Summary and conclusion}\label{sec:summary}                                                                                                            
 %================================================
In this work we consider the proposed long baseline experiment P2O with a 2595 km baseline
from the already existing accelerator complex at Protvino to the far detector situated at the 
site of KM3NeT/ ORCA with a fiducial mass of approximately 6 Mt.  
and a peak energy of around 4-5 GeV. 
Such a long baseline offers very high sensitivity to neutrino mass hierarchy and the massive 
far detector provides very high statistics even with a relatively moderate 90 kW proton beam. 
In this work, we have discussed the capability of an LBL experiment to probe fundamental theories of 
quantum gravity that can potentially manifest itself in the form of Lorentz invariance violation (LIV) around this energy range. 
We first discuss how the probabilities can deviate from the standard interaction (SI) scenario by different Lorentz invariance violating parameters (which are also CPT-violating) 
at the P2O baseline. 
We then analytically derive the approximate changes in the appearance probabilities, $\dpme$ ($ = P_{\mu e} (\text{SI+LIV}) - P_{\mu e} (\text{SI})$)  that are induced 
by individual LIV parameters. 
We illustrate by means of heatplots of $\dpme$, how the LIV parameters impact at the 
baseline of P2O at its peak energy and compare it with DUNE experiment. 
In presence of $\aem$ and $\aet$, we find interesting degenerate branches (existing even for larger values of the LIV parameters) at specific values of the standard CP phase $\dcp$. 
As a function of $\tc$, we observe that the impact of $\aem$ on $\dpme$ is slightly higher 
than that of $\aet$. 
These features were also explained with the help of probability expressions in presence of these two parameters.
We also find two degenerate regions at the level of probability for $\aee \simeq 0, 22 \times 10^{-23}$ GeV, 
whereas for DUNE we find only the trivial one at $\aee = 0$. 
We explain this by breaking down the corresponding $\dpme$ and showing that the relevant 
{\textit{sine}}-term oscillates faster for P2O (due to higher peak energy and a slightly higher value of average baseline density), - forcing a second nontrivial solution. 
We then proceed to estimate $\chisq$ sensitivities to LIV parameters for P2O alone and also discuss  
how significantly the results improve when the simulated data of P2O is combined with that of  DUNE. 
For completeness, we have also compared the results with a similar analysis using 
the simulated data of only DUNE.
 The sensitivity analyses were carried out by showing correlations of the LIV parameters ($\aee, \amm, \aem, \aet, \amt$) 
 among themselves and also with the two standard oscillation parameters $\dcp$ and $\tc$. 
 For $\aee$ we discuss in detail analytically how a crucial change in sign due to marginalization over the opposite mass hierarchy produces a {\textit{fake}} $\chisq$ minimum  
 around $\aee \simeq -22 \times 10^{-23}$ GeV. 
 For all the parameter spaces we numerically estimate the area of the regions  that are excluded (at $95\%$ C.L.) by P2O (or DUNE) individually and compare it to that by the combined 
 analysis of (P2O + DUNE). 
 The significance quantitative increase in the excluded area for (P2O+DUNE) shows the overwhelming advantage of the combined analysis in all cases. 
 Finally we calculate the one-dimensional $\chisq$ projections as a function of all five individual LIV parameters after marginalisation over all other parameters and estimate the $95\%$ C.L. constraints. 
% On comparison with similar analyses done using DUNE alone, 
 We find that  
 for the diagonal LIV parameters there is a significant improvement of the constraints estimated with the combined (P2O+DUNE) analysis. 
 Especially noteworthy is the lifting of degeneracy around $\aee \simeq -22 \times 10^{-23}$ GeV, which is not possible with P2O-only or DUNE-only analysis. 
 For the off-diagonal LIV parameters also our estimated bounds improve moderately.    
%==============================================
 \section*{Acknowledgement}                                                                                                            
 %================================================
MM acknowledges the support from IBS under the project code IBS-R018-D1. 
We thank D. Zabarov for providing us with the P2O flux files.
NRKC gratefully acknowledges the financial support of the Ministry of Science, Innovation and
Universities: State Program of Generation of Knowledge, ref. PGC2018-096663-B-C41 (MCIU /
FEDER), Spain.
NF is grateful for a visiting position at IBS CTPU where this work was finished.
We would also like to thank the anonymous referee for the valuable suggestions.

%==============================================
 \bibliographystyle{JHEP}
\bibliography{reference}

\providecommand{\href}[2]{#2}\begingroup\raggedright\begin{thebibliography}{10}

\bibitem{Fukuda:1998mi}
{\scshape Super-Kamiokande Collaboration} collaboration, \emph{{Evidence for
  oscillation of atmospheric neutrinos}},
  \href{https://doi.org/10.1103/PhysRevLett.81.1562}{\emph{Phys.Rev.Lett.}
  {\bfseries 81} (1998) 1562}
  [\href{https://arxiv.org/abs/hep-ex/9807003}{{\ttfamily hep-ex/9807003}}].

\bibitem{Ahmad:2002jz}
{\scshape SNO} collaboration, \emph{{Direct evidence for neutrino flavor
  transformation from neutral current interactions in the Sudbury Neutrino
  Observatory}},
  \href{https://doi.org/10.1103/PhysRevLett.89.011301}{\emph{Phys. Rev. Lett.}
  {\bfseries 89} (2002) 011301}
  [\href{https://arxiv.org/abs/nucl-ex/0204008}{{\ttfamily nucl-ex/0204008}}].

\bibitem{Sakharov:1967dj}
A.~Sakharov, \emph{{Violation of CP Invariance, C asymmetry, and baryon
  asymmetry of the universe}},
  \href{https://doi.org/10.1070/PU1991v034n05ABEH002497}{\emph{Sov. Phys. Usp.}
  {\bfseries 34} (1991) 392}.

\bibitem{Abe:2013hdq}
{\scshape T2K} collaboration, \emph{{Observation of Electron Neutrino
  Appearance in a Muon Neutrino Beam}},
  \href{https://doi.org/10.1103/PhysRevLett.112.061802}{\emph{Phys. Rev. Lett.}
  {\bfseries 112} (2014) 061802}
  [\href{https://arxiv.org/abs/1311.4750}{{\ttfamily 1311.4750}}].

\bibitem{Ayres:2004js}
{\scshape NOvA} collaboration, \emph{{NOvA: Proposal to Build a 30 Kiloton
  Off-Axis Detector to Study $\nu_{\mu} \to \nu_e$ Oscillations in the NuMI
  Beamline}},  \href{https://arxiv.org/abs/hep-ex/0503053}{{\ttfamily
  hep-ex/0503053}}.

\bibitem{Abe:2019vii}
{\scshape T2K} collaboration, \emph{{Constraint on the matter--antimatter
  symmetry-violating phase in neutrino oscillations}},
  \href{https://doi.org/10.1038/s41586-020-2177-0}{\emph{Nature} {\bfseries
  580} (2020) 339} [\href{https://arxiv.org/abs/1910.03887}{{\ttfamily
  1910.03887}}].

\bibitem{Acero:2019ksn}
{\scshape NOvA} collaboration, \emph{{First Measurement of Neutrino Oscillation
  Parameters using Neutrinos and Antineutrinos by NOvA}},
  \href{https://doi.org/10.1103/PhysRevLett.123.151803}{\emph{Phys. Rev. Lett.}
  {\bfseries 123} (2019) 151803}
  [\href{https://arxiv.org/abs/1906.04907}{{\ttfamily 1906.04907}}].

\bibitem{deSalas:2020pgw}
P.~de~Salas, D.~Forero, S.~Gariazzo, P.~Mart{\'i}nez-Mirav{\'e}, O.~Mena,
  C.~Ternes et~al., \emph{{2020 Global reassessment of the neutrino oscillation
  picture}},  \href{https://arxiv.org/abs/2006.11237}{{\ttfamily 2006.11237}}.

\bibitem{globalfit}
Valencia-Globalfit. \url{http://globalfit.astroparticles.es/}, 2020.

\bibitem{Capozzi:2017ipn}
F.~Capozzi, E.~Di~Valentino, E.~Lisi, A.~Marrone, A.~Melchiorri and A.~Palazzo,
  \emph{{Global constraints on absolute neutrino masses and their ordering}},
  \href{https://doi.org/10.1103/PhysRevD.95.096014}{\emph{Phys. Rev. D}
  {\bfseries 95} (2017) 096014}
  [\href{https://arxiv.org/abs/2003.08511}{{\ttfamily 2003.08511}}].

\bibitem{Esteban:2018azc}
I.~Esteban, M.C.~Gonzalez-Garcia, A.~Hernandez-Cabezudo, M.~Maltoni and
  T.~Schwetz, \emph{{Global analysis of three-flavour neutrino oscillations:
  synergies and tensions in the determination of $\theta_{23}$, $\delta_{CP}$,
  and the mass ordering}},
  \href{https://doi.org/10.1007/JHEP01(2019)106}{\emph{JHEP} {\bfseries 01}
  (2019) 106} [\href{https://arxiv.org/abs/1811.05487}{{\ttfamily
  1811.05487}}].

\bibitem{Acciarri:2015uup}
{\scshape DUNE} collaboration, \emph{{Long-Baseline Neutrino Facility (LBNF)
  and Deep Underground Neutrino Experiment (DUNE) Conceptual Design Report
  Volume 2: The Physics Program for DUNE at LBNF}},
  \href{https://arxiv.org/abs/1512.06148}{{\ttfamily 1512.06148}}.

\bibitem{Abi:2020evt}
{\scshape DUNE} collaboration, \emph{{Deep Underground Neutrino Experiment
  (DUNE), Far Detector Technical Design Report, Volume II DUNE Physics}},
  \href{https://arxiv.org/abs/2002.03005}{{\ttfamily 2002.03005}}.

\bibitem{Abe:2015zbg}
{\scshape Hyper-Kamiokande Proto-Collaboration} collaboration, \emph{{Physics
  potential of a long-baseline neutrino oscillation experiment using a J-PARC
  neutrino beam and Hyper-Kamiokande}},
  \href{https://doi.org/10.1093/ptep/ptv061}{\emph{PTEP} {\bfseries 2015}
  (2015) 053C02} [\href{https://arxiv.org/abs/1502.05199}{{\ttfamily
  1502.05199}}].

\bibitem{Abe:2016ero}
{\scshape Hyper-Kamiokande} collaboration, \emph{{Physics potentials with the
  second Hyper-Kamiokande detector in Korea}},
  \href{https://doi.org/10.1093/ptep/pty044}{\emph{PTEP} {\bfseries 2018}
  (2018) 063C01} [\href{https://arxiv.org/abs/1611.06118}{{\ttfamily
  1611.06118}}].

\bibitem{Baussan:2013zcy}
{\scshape ESSnuSB} collaboration, \emph{{A very intense neutrino super beam
  experiment for leptonic CP violation discovery based on the European
  spallation source linac}},
  \href{https://doi.org/10.1016/j.nuclphysb.2014.05.016}{\emph{Nucl. Phys. B}
  {\bfseries 885} (2014) 127}
  [\href{https://arxiv.org/abs/1309.7022}{{\ttfamily 1309.7022}}].

\bibitem{JUNO:2015zny}
{\scshape JUNO} collaboration, \emph{{Neutrino Physics with JUNO}},
  \href{https://doi.org/10.1088/0954-3899/43/3/030401}{\emph{J. Phys. G}
  {\bfseries 43} (2016) 030401}
  [\href{https://arxiv.org/abs/1507.05613}{{\ttfamily 1507.05613}}].

\bibitem{Akindinov:2019flp}
A.V.~Akindinov et~al., \emph{{Letter of Interest for a Neutrino Beam from
  Protvino to KM3NeT/ORCA}},
  \href{https://doi.org/10.1140/epjc/s10052-019-7259-5}{\emph{Eur. Phys. J. C}
  {\bfseries 79} (2019) 758}
  [\href{https://arxiv.org/abs/1902.06083}{{\ttfamily 1902.06083}}].

\bibitem{Greenberg:2002uu}
O.W.~Greenberg, \emph{{CPT violation implies violation of Lorentz invariance}},
  \href{https://doi.org/10.1103/PhysRevLett.89.231602}{\emph{Phys. Rev. Lett.}
  {\bfseries 89} (2002) 231602}
  [\href{https://arxiv.org/abs/hep-ph/0201258}{{\ttfamily hep-ph/0201258}}].

\bibitem{Kostelecky:1988zi}
V.A.~Kostelecky and S.~Samuel, \emph{{Spontaneous Breaking of Lorentz Symmetry
  in String Theory}},
  \href{https://doi.org/10.1103/PhysRevD.39.683}{\emph{Phys. Rev.} {\bfseries
  D39} (1989) 683}.

\bibitem{Kostelecky:1989jp}
V.A.~Kostelecky and S.~Samuel, \emph{{Phenomenological Gravitational
  Constraints on Strings and Higher Dimensional Theories}},
  \href{https://doi.org/10.1103/PhysRevLett.63.224}{\emph{Phys. Rev. Lett.}
  {\bfseries 63} (1989) 224}.

\bibitem{Kostelecky:1991ak}
V.A.~Kostelecky and R.~Potting, \emph{{CPT and strings}},
  \href{https://doi.org/10.1016/0550-3213(91)90071-5}{\emph{Nucl. Phys.}
  {\bfseries B359} (1991) 545}.

\bibitem{Kostelecky:1994rn}
V.A.~Kostelecky and R.~Potting, \emph{{CPT, strings, and meson factories}},
  \href{https://doi.org/10.1103/PhysRevD.51.3923}{\emph{Phys. Rev.} {\bfseries
  D51} (1995) 3923} [\href{https://arxiv.org/abs/hep-ph/9501341}{{\ttfamily
  hep-ph/9501341}}].

\bibitem{Kostelecky:1995qk}
V.A.~Kostelecky and R.~Potting, \emph{{Expectation values, Lorentz invariance,
  and CPT in the open bosonic string}},
  \href{https://doi.org/10.1016/0370-2693(96)00589-8}{\emph{Phys. Lett.}
  {\bfseries B381} (1996) 89}
  [\href{https://arxiv.org/abs/hep-th/9605088}{{\ttfamily hep-th/9605088}}].

\bibitem{Colladay:1996iz}
D.~Colladay and V.A.~Kostelecky, \emph{{CPT violation and the standard model}},
  \href{https://doi.org/10.1103/PhysRevD.55.6760}{\emph{Phys. Rev.} {\bfseries
  D55} (1997) 6760} [\href{https://arxiv.org/abs/hep-ph/9703464}{{\ttfamily
  hep-ph/9703464}}].

\bibitem{Colladay:1998fq}
D.~Colladay and V.A.~Kostelecky, \emph{{Lorentz violating extension of the
  standard model}},
  \href{https://doi.org/10.1103/PhysRevD.58.116002}{\emph{Phys. Rev.}
  {\bfseries D58} (1998) 116002}
  [\href{https://arxiv.org/abs/hep-ph/9809521}{{\ttfamily hep-ph/9809521}}].

\bibitem{Kostelecky:2003fs}
V.A.~Kostelecky, \emph{{Gravity, Lorentz violation, and the standard model}},
  \href{https://doi.org/10.1103/PhysRevD.69.105009}{\emph{Phys. Rev. D}
  {\bfseries 69} (2004) 105009}
  [\href{https://arxiv.org/abs/hep-th/0312310}{{\ttfamily hep-th/0312310}}].

\bibitem{LSND:2005oop}
{\scshape LSND} collaboration, \emph{{Tests of Lorentz violation in anti-nu(mu)
  ---\ensuremath{>} anti-nu(e) oscillations}},
  \href{https://doi.org/10.1103/PhysRevD.72.076004}{\emph{Phys. Rev. D}
  {\bfseries 72} (2005) 076004}
  [\href{https://arxiv.org/abs/hep-ex/0506067}{{\ttfamily hep-ex/0506067}}].

\bibitem{MINOS:2008fnv}
{\scshape MINOS} collaboration, \emph{{Testing Lorentz Invariance and CPT
  Conservation with NuMI Neutrinos in the MINOS Near Detector}},
  \href{https://doi.org/10.1103/PhysRevLett.101.151601}{\emph{Phys. Rev. Lett.}
  {\bfseries 101} (2008) 151601}
  [\href{https://arxiv.org/abs/0806.4945}{{\ttfamily 0806.4945}}].

\bibitem{MINOS:2010kat}
{\scshape MINOS} collaboration, \emph{{A Search for Lorentz Invariance and CPT
  Violation with the MINOS Far Detector}},
  \href{https://doi.org/10.1103/PhysRevLett.105.151601}{\emph{Phys. Rev. Lett.}
  {\bfseries 105} (2010) 151601}
  [\href{https://arxiv.org/abs/1007.2791}{{\ttfamily 1007.2791}}].

\bibitem{MiniBooNE:2011pix}
{\scshape MiniBooNE} collaboration, \emph{{Test of Lorentz and CPT violation
  with Short Baseline Neutrino Oscillation Excesses}},
  \href{https://doi.org/10.1016/j.physletb.2012.12.020}{\emph{Phys. Lett. B}
  {\bfseries 718} (2013) 1303}
  [\href{https://arxiv.org/abs/1109.3480}{{\ttfamily 1109.3480}}].

\bibitem{DoubleChooz:2012eiq}
{\scshape Double Chooz} collaboration, \emph{{First Test of Lorentz Violation
  with a Reactor-based Antineutrino Experiment}},
  \href{https://doi.org/10.1103/PhysRevD.86.112009}{\emph{Phys. Rev. D}
  {\bfseries 86} (2012) 112009}
  [\href{https://arxiv.org/abs/1209.5810}{{\ttfamily 1209.5810}}].

\bibitem{Super-Kamiokande:2014exs}
{\scshape Super-Kamiokande} collaboration, \emph{{Test of Lorentz invariance
  with atmospheric neutrinos}},
  \href{https://doi.org/10.1103/PhysRevD.91.052003}{\emph{Phys. Rev. D}
  {\bfseries 91} (2015) 052003}
  [\href{https://arxiv.org/abs/1410.4267}{{\ttfamily 1410.4267}}].

\bibitem{T2K:2017ega}
{\scshape T2K} collaboration, \emph{{Search for Lorentz and CPT violation using
  sidereal time dependence of neutrino flavor transitions over a short
  baseline}}, \href{https://doi.org/10.1103/PhysRevD.95.111101}{\emph{Phys.
  Rev. D} {\bfseries 95} (2017) 111101}
  [\href{https://arxiv.org/abs/1703.01361}{{\ttfamily 1703.01361}}].

\bibitem{IceCube:2017qyp}
{\scshape IceCube} collaboration, \emph{{Neutrino Interferometry for
  High-Precision Tests of Lorentz Symmetry with IceCube}},
  \href{https://doi.org/10.1038/s41567-018-0172-2}{\emph{Nature Phys.}
  {\bfseries 14} (2018) 961}
  [\href{https://arxiv.org/abs/1709.03434}{{\ttfamily 1709.03434}}].

\bibitem{Dighe:2008bu}
A.~Dighe and S.~Ray, \emph{{CPT violation in long baseline neutrino
  experiments: A Three flavor analysis}},
  \href{https://doi.org/10.1103/PhysRevD.78.036002}{\emph{Phys.Rev.} {\bfseries
  D78} (2008) 036002} [\href{https://arxiv.org/abs/0802.0121}{{\ttfamily
  0802.0121}}].

\bibitem{Barenboim:2009ts}
G.~Barenboim and J.D.~Lykken, \emph{{MINOS and CPT-violating neutrinos}},
  \href{https://doi.org/10.1103/PhysRevD.80.113008}{\emph{Phys. Rev.}
  {\bfseries D80} (2009) 113008}
  [\href{https://arxiv.org/abs/0908.2993}{{\ttfamily 0908.2993}}].

\bibitem{Rebel:2013vc}
B.~Rebel and S.~Mufson, \emph{{The Search for Neutrino-Antineutrino Mixing
  Resulting from Lorentz Invariance Violation using neutrino interactions in
  MINOS}},
  \href{https://doi.org/10.1016/j.astropartphys.2013.07.006}{\emph{Astropart.
  Phys.} {\bfseries 48} (2013) 78}
  [\href{https://arxiv.org/abs/1301.4684}{{\ttfamily 1301.4684}}].

\bibitem{deGouvea:2017yvn}
A.~de~Gouv\^ea and K.J.~Kelly, \emph{{Neutrino vs. Antineutrino Oscillation
  Parameters at DUNE and Hyper-Kamiokande}},
  \href{https://doi.org/10.1103/PhysRevD.96.095018}{\emph{Phys. Rev. D}
  {\bfseries 96} (2017) 095018}
  [\href{https://arxiv.org/abs/1709.06090}{{\ttfamily 1709.06090}}].

\bibitem{Barenboim:2017ewj}
G.~Barenboim, C.A.~Ternes and M.~T{\'o}rtola, \emph{{Neutrinos, DUNE and the
  world best bound on CPT violation}},
  \href{https://arxiv.org/abs/1712.01714}{{\ttfamily 1712.01714}}.

\bibitem{Barenboim:2018ctx}
G.~Barenboim, M.~Masud, C.A.~Ternes and M.~T\'ortola, \emph{{Exploring the
  intrinsic Lorentz-violating parameters at DUNE}},
  \href{https://doi.org/10.1016/j.physletb.2018.11.040}{\emph{Phys. Lett. B}
  {\bfseries 788} (2019) 308}
  [\href{https://arxiv.org/abs/1805.11094}{{\ttfamily 1805.11094}}].

\bibitem{Majhi:2019tfi}
R.~Majhi, S.~Chembra and R.~Mohanta, \emph{{Exploring the effect of Lorentz
  invariance violation with the currently running long-baseline experiments}},
  \href{https://doi.org/10.1140/epjc/s10052-020-7963-1}{\emph{Eur. Phys. J. C}
  {\bfseries 80} (2020) 364}
  [\href{https://arxiv.org/abs/1907.09145}{{\ttfamily 1907.09145}}].

\bibitem{KumarAgarwalla:2019gdj}
S.~Kumar~Agarwalla and M.~Masud, \emph{{Can Lorentz invariance violation affect
  the sensitivity of deep underground neutrino experiment?}},
  \href{https://doi.org/10.1140/epjc/s10052-020-8303-1}{\emph{Eur. Phys. J. C}
  {\bfseries 80} (2020) 716}
  [\href{https://arxiv.org/abs/1912.13306}{{\ttfamily 1912.13306}}].

\bibitem{Rahaman:2021leu}
U.~Rahaman, \emph{{Looking for Lorentz invariance violation (LIV) in the latest
  long baseline accelerator neutrino oscillation data}},
  \href{https://doi.org/10.1140/epjc/s10052-021-09598-4}{\emph{Eur. Phys. J. C}
  {\bfseries 81} (2021) 792}
  [\href{https://arxiv.org/abs/2103.04576}{{\ttfamily 2103.04576}}].

\bibitem{Giunti:2010zs}
C.~Giunti and M.~Laveder, \emph{{Hint of CPT Violation in Short-Baseline
  Electron Neutrino Disappearance}},
  \href{https://doi.org/10.1103/PhysRevD.82.113009}{\emph{Phys. Rev.}
  {\bfseries D82} (2010) 113009}
  [\href{https://arxiv.org/abs/1008.4750}{{\ttfamily 1008.4750}}].

\bibitem{Datta:2003dg}
A.~Datta, R.~Gandhi, P.~Mehta and S.U.~Sankar, \emph{{Atmospheric neutrinos as
  a probe of CPT and Lorentz violation}},
  \href{https://doi.org/10.1016/j.physletb.2004.07.035}{\emph{Phys. Lett.}
  {\bfseries B597} (2004) 356}
  [\href{https://arxiv.org/abs/hep-ph/0312027}{{\ttfamily hep-ph/0312027}}].

\bibitem{Chatterjee:2014oda}
A.~Chatterjee, R.~Gandhi and J.~Singh, \emph{{Probing Lorentz and CPT Violation
  in a Magnetized Iron Detector using Atmospheric Neutrinos}},
  \href{https://doi.org/10.1007/JHEP06(2014)045}{\emph{JHEP} {\bfseries 1406}
  (2014) 045} [\href{https://arxiv.org/abs/1402.6265}{{\ttfamily 1402.6265}}].

\bibitem{Koranga:2014dua}
B.~Singh~Koranga and P.~Khurana, \emph{{CPT Violation in Atmospheric Neutrino
  Oscillation: A Two Flavour Matter Effects}},
  \href{https://doi.org/10.1007/s10773-014-2126-5}{\emph{Int. J. Theor. Phys.}
  {\bfseries 53} (2014) 3737}.

\bibitem{Sahoo:2021dit}
S.~Sahoo, A.~Kumar and S.K.~Agarwalla, \emph{{Probing Lorentz Invariance
  Violation with atmospheric neutrinos at INO-ICAL}},
  \href{https://doi.org/10.1007/JHEP03(2022)050}{\emph{JHEP} {\bfseries 03}
  (2022) 050} [\href{https://arxiv.org/abs/2110.13207}{{\ttfamily
  2110.13207}}].

\bibitem{Diaz:2016fqd}
J.S.~Diaz and T.~Schwetz, \emph{{Limits on CPT violation from solar
  neutrinos}}, \href{https://doi.org/10.1103/PhysRevD.93.093004}{\emph{Phys.
  Rev.} {\bfseries D93} (2016) 093004}
  [\href{https://arxiv.org/abs/1603.04468}{{\ttfamily 1603.04468}}].

\bibitem{Hooper:2005jp}
D.~Hooper, D.~Morgan and E.~Winstanley, \emph{{Lorentz and CPT invariance
  violation in high-energy neutrinos}},
  \href{https://doi.org/10.1103/PhysRevD.72.065009}{\emph{Phys. Rev.}
  {\bfseries D72} (2005) 065009}
  [\href{https://arxiv.org/abs/hep-ph/0506091}{{\ttfamily hep-ph/0506091}}].

\bibitem{Tomar:2015fha}
G.~Tomar, S.~Mohanty and S.~Pakvasa, \emph{{Lorentz Invariance Violation and
  IceCube Neutrino Events}},
  \href{https://doi.org/10.1007/JHEP11(2015)022}{\emph{JHEP} {\bfseries 11}
  (2015) 022} [\href{https://arxiv.org/abs/1507.03193}{{\ttfamily
  1507.03193}}].

\bibitem{Liao:2017yuy}
J.~Liao and D.~Marfatia, \emph{{IceCube?s astrophysical neutrino energy
  spectrum from CPT violation}},
  \href{https://doi.org/10.1103/PhysRevD.97.041302}{\emph{Phys. Rev.}
  {\bfseries D97} (2018) 041302}
  [\href{https://arxiv.org/abs/1711.09266}{{\ttfamily 1711.09266}}].

\bibitem{Lin:2021cst}
H.-X.~Lin, P.~Pasquini, J.~Tang and S.~Vihonen, \emph{{Nonminimal Lorentz
  invariance violation in light of the muon anomalous magnetic moment and
  long-baseline neutrino oscillation data}},
  \href{https://doi.org/10.1103/PhysRevD.105.096029}{\emph{Phys. Rev. D}
  {\bfseries 105} (2022) 096029}
  [\href{https://arxiv.org/abs/2111.14336}{{\ttfamily 2111.14336}}].

\bibitem{Kostelecky:2008ts}
V.A.~Kostelecky and N.~Russell, \emph{{Data Tables for Lorentz and CPT
  Violation}},  \href{https://arxiv.org/abs/0801.0287}{{\ttfamily 0801.0287}}.

\bibitem{zaborov_talk}
D.~Zaborov, ``{Scientific Potential of a neutrino beam from Protvino to ORCA
  (P2O)}.'' talk at Neutrino GDR Meeting, Paris, November 2017.
  "\url{https://indico.in2p3.fr/event/16553/contributions/57491/attachments/45237/56246/P2O-zaborov-GDR-neutrino-Nov2017.pdf}",
  2017.

\bibitem{KM3Net:2016zxf}
{\scshape KM3Net} collaboration, \emph{{Letter of intent for KM3NeT 2.0}},
  \href{https://doi.org/10.1088/0954-3899/43/8/084001}{\emph{J. Phys. G}
  {\bfseries 43} (2016) 084001}
  [\href{https://arxiv.org/abs/1601.07459}{{\ttfamily 1601.07459}}].

\bibitem{Zaborov:2018whl}
{\scshape KM3NeT} collaboration, \emph{{The KM3NeT Neutrino Telescope and the
  potential of a neutrino beam from Russia to the Mediterranean Sea}},  in
  \emph{{18th Lomonosov Conference on Elementary Particle Physics}},
  pp.~53--60, 2019, \href{https://doi.org/10.1142/9789811202339_0009}{DOI}
  [\href{https://arxiv.org/abs/1803.08017}{{\ttfamily 1803.08017}}].

\bibitem{Coloma:2013rqa}
P.~Coloma and P.~Huber, \emph{{Impact of nuclear effects on the extraction of
  neutrino oscillation parameters}},
  \href{https://doi.org/10.1103/PhysRevLett.111.221802}{\emph{Phys. Rev. Lett.}
  {\bfseries 111} (2013) 221802}
  [\href{https://arxiv.org/abs/1307.1243}{{\ttfamily 1307.1243}}].

\bibitem{Mosel:2013fxa}
U.~Mosel, O.~Lalakulich and K.~Gallmeister, \emph{{Energy reconstruction in the
  Long-Baseline Neutrino Experiment}},
  \href{https://doi.org/10.1103/PhysRevLett.112.151802}{\emph{Phys. Rev. Lett.}
  {\bfseries 112} (2014) 151802}
  [\href{https://arxiv.org/abs/1311.7288}{{\ttfamily 1311.7288}}].

\bibitem{Alvarez-Ruso:2014bla}
L.~Alvarez-Ruso, Y.~Hayato and J.~Nieves, \emph{{Progress and open questions in
  the physics of neutrino cross sections at intermediate energies}},
  \href{https://doi.org/10.1088/1367-2630/16/7/075015}{\emph{New J. Phys.}
  {\bfseries 16} (2014) 075015}
  [\href{https://arxiv.org/abs/1403.2673}{{\ttfamily 1403.2673}}].

\bibitem{Benhar:2015wva}
O.~Benhar, P.~Huber, C.~Mariani and D.~Meloni,
  \emph{{Neutrino\textendash{}nucleus interactions and the determination of
  oscillation parameters}},
  \href{https://doi.org/10.1016/j.physrep.2017.07.004}{\emph{Phys. Rept.}
  {\bfseries 700} (2017) 1} [\href{https://arxiv.org/abs/1501.06448}{{\ttfamily
  1501.06448}}].

\bibitem{NuSTEC:2017hzk}
{\scshape NuSTEC} collaboration, \emph{{NuSTEC White Paper: Status and
  challenges of neutrino\textendash{}nucleus scattering}},
  \href{https://doi.org/10.1016/j.ppnp.2018.01.006}{\emph{Prog. Part. Nucl.
  Phys.} {\bfseries 100} (2018) 1}
  [\href{https://arxiv.org/abs/1706.03621}{{\ttfamily 1706.03621}}].

\bibitem{Nagu:2019fvi}
S.~Nagu, J.~Singh and J.~Singh, \emph{{Nuclear Effects and CP Sensitivity at
  DUNE}}, \href{https://doi.org/10.1155/2020/5472713}{\emph{Adv. High Energy
  Phys.} {\bfseries 2020} (2020) 5472713}
  [\href{https://arxiv.org/abs/1906.02190}{{\ttfamily 1906.02190}}].

\bibitem{Singha:2021jkn}
D.K.~Singha, M.~Ghosh, R.~Majhi and R.~Mohanta, \emph{{Optimal configuration of
  Protvino to ORCA experiment for hierarchy and non-standard interactions}},
  \href{https://doi.org/10.1007/JHEP05(2022)117}{\emph{JHEP} {\bfseries 05}
  (2022) 117} [\href{https://arxiv.org/abs/2112.04876}{{\ttfamily
  2112.04876}}].

\bibitem{Choubey:2018rnl}
S.~Choubey, M.~Ghosh and D.~Pramanik, \emph{{Sensitivity study of Protvino to
  ORCA (P2O) experiment: effect of antineutrino run, background and
  systematics}},
  \href{https://doi.org/10.1140/epjc/s10052-019-7064-1}{\emph{Eur. Phys. J. C}
  {\bfseries 79} (2019) 603}
  [\href{https://arxiv.org/abs/1812.02608}{{\ttfamily 1812.02608}}].

\bibitem{Perrin-Terrin:2021jtl}
M.~Perrin-Terrin, \emph{{Neutrino tagging: a new tool for accelerator based
  neutrino experiments}},
  \href{https://doi.org/10.1140/epjc/s10052-022-10397-8}{\emph{Eur. Phys. J. C}
  {\bfseries 82} (2022) 465}
  [\href{https://arxiv.org/abs/2112.12848}{{\ttfamily 2112.12848}}].

\bibitem{Kaur:2021rau}
D.~Kaur, N.R.K.~Chowdhury and U.~Rahaman, \emph{{Effect of non-unitary mixing
  on the mass hierarchy and CP violation determination at the Protvino to Orca
  experiment}},  \href{https://arxiv.org/abs/2110.02917}{{\ttfamily
  2110.02917}}.

\bibitem{Kostelecky:2000mm}
V.A.~Kostelecky and R.~Lehnert, \emph{{Stability, causality, and Lorentz and
  CPT violation}},
  \href{https://doi.org/10.1103/PhysRevD.63.065008}{\emph{Phys. Rev. D}
  {\bfseries 63} (2001) 065008}
  [\href{https://arxiv.org/abs/hep-th/0012060}{{\ttfamily hep-th/0012060}}].

\bibitem{Kostelecky:2003cr}
V.A.~Kostelecky and M.~Mewes, \emph{{Lorentz and CPT violation in neutrinos}},
  \href{https://doi.org/10.1103/PhysRevD.69.016005}{\emph{Phys. Rev.}
  {\bfseries D69} (2004) 016005}
  [\href{https://arxiv.org/abs/hep-ph/0309025}{{\ttfamily hep-ph/0309025}}].

\bibitem{Diaz:2009qk}
J.S.~Diaz, V.A.~Kostelecky and M.~Mewes, \emph{{Perturbative Lorentz and CPT
  violation for neutrino and antineutrino oscillations}},
  \href{https://doi.org/10.1103/PhysRevD.80.076007}{\emph{Phys. Rev. D}
  {\bfseries 80} (2009) 076007}
  [\href{https://arxiv.org/abs/0908.1401}{{\ttfamily 0908.1401}}].

\bibitem{Kostelecky:2011gq}
A.~Kostelecky and M.~Mewes, \emph{{Neutrinos with Lorentz-violating operators
  of arbitrary dimension}},
  \href{https://doi.org/10.1103/PhysRevD.85.096005}{\emph{Phys. Rev.}
  {\bfseries D85} (2012) 096005}
  [\href{https://arxiv.org/abs/1112.6395}{{\ttfamily 1112.6395}}].

\bibitem{Kostelecky:2002hh}
V.A.~Kostelecky and M.~Mewes, \emph{{Signals for Lorentz violation in
  electrodynamics}},
  \href{https://doi.org/10.1103/PhysRevD.66.056005}{\emph{Phys. Rev. D}
  {\bfseries 66} (2002) 056005}
  [\href{https://arxiv.org/abs/hep-ph/0205211}{{\ttfamily hep-ph/0205211}}].

\bibitem{Diaz:2011ia}
J.S.~Diaz and A.~Kostelecky, \emph{{Lorentz- and CPT-violating models for
  neutrino oscillations}},
  \href{https://doi.org/10.1103/PhysRevD.85.016013}{\emph{Phys. Rev. D}
  {\bfseries 85} (2012) 016013}
  [\href{https://arxiv.org/abs/1108.1799}{{\ttfamily 1108.1799}}].

\bibitem{Diaz:2015dxa}
J.S.~Diaz, \emph{{Correspondence between nonstandard interactions and CPT
  violation in neutrino oscillations}},
  \href{https://arxiv.org/abs/1506.01936}{{\ttfamily 1506.01936}}.

\bibitem{Kikuchi:2008vq}
T.~Kikuchi, H.~Minakata and S.~Uchinami, \emph{{Perturbation Theory of Neutrino
  Oscillation with Nonstandard Neutrino Interactions}},
  \href{https://doi.org/10.1088/1126-6708/2009/03/114}{\emph{JHEP} {\bfseries
  0903} (2009) 114} [\href{https://arxiv.org/abs/0809.3312}{{\ttfamily
  0809.3312}}].

\bibitem{Agarwalla:2016fkh}
S.K.~Agarwalla, S.S.~Chatterjee and A.~Palazzo, \emph{{Degeneracy between
  $\theta_{23}$ octant and neutrino non-standard interactions at DUNE}},
  \href{https://doi.org/10.1016/j.physletb.2016.09.020}{\emph{Phys. Lett. B}
  {\bfseries 762} (2016) 64}
  [\href{https://arxiv.org/abs/1607.01745}{{\ttfamily 1607.01745}}].

\bibitem{Masud:2018pig}
M.~Masud, S.~Roy and P.~Mehta, \emph{{Correlations and degeneracies among the
  NSI parameters with tunable beams at DUNE}},
  \href{https://doi.org/10.1103/PhysRevD.99.115032}{\emph{Phys. Rev. D}
  {\bfseries 99} (2019) 115032}
  [\href{https://arxiv.org/abs/1812.10290}{{\ttfamily 1812.10290}}].

\bibitem{Huber:2004ka}
P.~Huber, M.~Lindner and W.~Winter, \emph{{Simulation of long-baseline neutrino
  oscillation experiments with GLoBES (General Long Baseline Experiment
  Simulator)}}, \href{https://doi.org/10.1016/j.cpc.2005.01.003}{\emph{Comput.
  Phys. Commun.} {\bfseries 167} (2005) 195}
  [\href{https://arxiv.org/abs/hep-ph/0407333}{{\ttfamily hep-ph/0407333}}].

\bibitem{Huber:2007ji}
P.~Huber, J.~Kopp, M.~Lindner, M.~Rolinec and W.~Winter, \emph{{New features in
  the simulation of neutrino oscillation experiments with GLoBES 3.0: General
  Long Baseline Experiment Simulator}},
  \href{https://doi.org/10.1016/j.cpc.2007.05.004}{\emph{Comput. Phys. Commun.}
  {\bfseries 177} (2007) 432}
  [\href{https://arxiv.org/abs/hep-ph/0701187}{{\ttfamily hep-ph/0701187}}].

\bibitem{Kopp:2006wp}
J.~Kopp, \emph{{Efficient numerical diagonalization of hermitian 3 x 3
  matrices}}, \href{https://doi.org/10.1142/S0129183108012303}{\emph{Int. J.
  Mod. Phys.} {\bfseries C19} (2008) 523}
  [\href{https://arxiv.org/abs/physics/0610206}{{\ttfamily physics/0610206}}].

\bibitem{Kopp:2007ne}
J.~Kopp, M.~Lindner, T.~Ota and J.~Sato, \emph{{Non-standard neutrino
  interactions in reactor and superbeam experiments}},
  \href{https://doi.org/10.1103/PhysRevD.77.013007}{\emph{Phys. Rev.}
  {\bfseries D77} (2008) 013007}
  [\href{https://arxiv.org/abs/0708.0152}{{\ttfamily 0708.0152}}].

\bibitem{Alion:2016uaj}
{\scshape DUNE} collaboration, \emph{{Experiment Simulation Configurations Used
  in DUNE CDR}},  \href{https://arxiv.org/abs/1606.09550}{{\ttfamily
  1606.09550}}.

\bibitem{Adrian-Martinez:2016fdl}
{\scshape KM3Net} collaboration, \emph{{Letter of Intent for KM3NeT2.0}},
  \href{https://arxiv.org/abs/1601.07459}{{\ttfamily 1601.07459}}.

\bibitem{Margiotta:2022kid}
{\scshape KM3NeT} collaboration, \emph{{The KM3NeT infrastructure: status and
  first results}},  in \emph{{21st International Symposium on Very High Energy
  Cosmic Ray Interactions}}, 8, 2022
  [\href{https://arxiv.org/abs/2208.07370}{{\ttfamily 2208.07370}}].

\bibitem{brunner_eppsu}
{\scshape P2O} collaboration, J.~Brunner, ``{Neutrino Beam from Protvino to
  KM3NeT/ORCA}.'' contribution at European Strategy for Particle Physics
  (2018-2020).
  "\url{https://indico.cern.ch/event/765096/contributions/3295791/attachments/1785302/2906340/Addendum_P2O.pdf}",
  2018.

\bibitem{p2o_brunner_talk_2021}
{\scshape P2O} collaboration, J.~Brunner, ``{P2O Status and current results}.''
  talk at P2O longbaseline Project, January, 2021.
  "\url{https://indico.cern.ch/event/997165/contributions/4189633/attachments/2175114/3672699/Intro_210121.pdf}",
  2021.

\bibitem{Huber:2002mx}
P.~Huber, M.~Lindner and W.~Winter, \emph{{Superbeams versus neutrino
  factories}}, \href{https://doi.org/10.1016/S0550-3213(02)00825-8}{\emph{Nucl.
  Phys.} {\bfseries B645} (2002) 3}
  [\href{https://arxiv.org/abs/hep-ph/0204352}{{\ttfamily hep-ph/0204352}}].

\bibitem{Fogli:2002pt}
G.L.~Fogli, E.~Lisi, A.~Marrone, D.~Montanino and A.~Palazzo, \emph{{Getting
  the most from the statistical analysis of solar neutrino oscillations}},
  \href{https://doi.org/10.1103/PhysRevD.66.053010}{\emph{Phys. Rev.}
  {\bfseries D66} (2002) 053010}
  [\href{https://arxiv.org/abs/hep-ph/0206162}{{\ttfamily hep-ph/0206162}}].

\bibitem{GonzalezGarcia:2004wg}
M.~Gonzalez-Garcia and M.~Maltoni, \emph{{Atmospheric neutrino oscillations and
  new physics}},
  \href{https://doi.org/10.1103/PhysRevD.70.033010}{\emph{Phys.Rev.} {\bfseries
  D70} (2004) 033010} [\href{https://arxiv.org/abs/hep-ph/0404085}{{\ttfamily
  hep-ph/0404085}}].

\bibitem{Gandhi:2007td}
R.~Gandhi, P.~Ghoshal, S.~Goswami, P.~Mehta, S.U.~Sankar and S.~Shalgar,
  \emph{{Mass Hierarchy Determination via future Atmospheric Neutrino
  Detectors}}, \href{https://doi.org/10.1103/PhysRevD.76.073012}{\emph{Phys.
  Rev.} {\bfseries D76} (2007) 073012}
  [\href{https://arxiv.org/abs/0707.1723}{{\ttfamily 0707.1723}}].

\bibitem{Qian:2012zn}
X.~Qian, A.~Tan, W.~Wang, J.J.~Ling, R.D.~McKeown and C.~Zhang,
  \emph{{Statistical Evaluation of Experimental Determinations of Neutrino Mass
  Hierarchy}}, \href{https://doi.org/10.1103/PhysRevD.86.113011}{\emph{Phys.
  Rev.} {\bfseries D86} (2012) 113011}
  [\href{https://arxiv.org/abs/1210.3651}{{\ttfamily 1210.3651}}].

\bibitem{IceCube:2014gqr}
{\scshape IceCube} collaboration, \emph{{IceCube-Gen2: A Vision for the Future
  of Neutrino Astronomy in Antarctica}},
  \href{https://arxiv.org/abs/1412.5106}{{\ttfamily 1412.5106}}.

\end{thebibliography}\endgroup
\end{document}